\documentclass[12pt,preprint]{aastex}

\slugcomment{Accepted for Publication in the {\it Astronomical Journal} on 10/18/2005}

\begin{document}

\def\nsnrinsurvey{95}
\def\nsnrdetected{18}
\def\nsnrconfused{17}
\def\nspechtwo{4}
\def\nspecpah{3}
\def\nspecion{4}
\def\nspecHOT{7}

\def\nusplussaken{4}
\def\nusarendt{6}
\def\nusarendtsaken{4}

\def\nmol{9}
\def\nion{3}
\def\npah{4}
\def\nmix{6}
\def\nspot{23}

\def\radec #1;#2;#3;#4;#5;#6 {{#1}$^{\rm h}${#2}$^{\rm m}${#3}$^{\rm s}${#4}$^\circ${#5}$'${#6}$''$}
\def\iraccolor #1;#2;#3;#4 {{#1}/{#2}/{#3}/{#4}}

\title{A {\it Spitzer Space Telescope}
Infrared Survey of Supernova Remnants in the Inner Galaxy}

\author{William T. Reach, Jeonghee Rho, Achim Tappe, Thomas G. Pannuti}

\affil{Spitzer Science Infrared Center, 
California Institute of Technology,
Pasadena, CA 91125}

\author{Crystal L. Brogan}
\affil{Institute for Astronomy, University of Hawaii,
640 N. A'ohoku Place, Hilo, HI 96720}

\author{Edward B. Churchwell, Marilyn R. Meade, Brian Babler}
\affil{Department of Astronomy, University of Wisconsin, 475 N. Charter Street, 
Madison, WI 53706}

\author{R\'emy Indebetouw}
\affil{Astronomy Department, University of Virginia, Charlottesville, VA 22904}

\author{Barbara A. Whitney}
\affil{Space Science Institute, Boulder, CO 80303}

\email{reach@ipac.caltech.edu}

\begin{abstract}
Using Infrared Array Camera (IRAC) images at 3.6, 4.5, 5.8, and 8 $\mu$m from 
the GLIMPSE Legacy science program on the {\it Spitzer} Space Telescope, 
we searched for
infrared counterparts to the \nsnrinsurvey\ 
known supernova remnants that are located within 
galactic longitudes $65^\circ>|l|>10^\circ$ and latitudes $|b|<1^\circ$.
Eighteen infrared counterparts were detected. Many other supernova
remnants could have significant infrared emission but are in portions of the Milky
Way too confused to allow separation from bright \ion{H}{2} regions and
pervasive mid-infrared emission from atomic and molecular clouds along
the line of sight. Infrared emission from supernova remnants originates
from synchrotron emission, shock-heated dust, atomic fine-structure lines,
and molecular lines.
The detected remnants are G11.2-0.3, Kes~69, G22.7-0.2, 3C~391, W~44, 3C~396,
3C~397, W~49B, G54.4-0.3, Kes~17, Kes~20A, RCW~103, G344.7-0.1, 
G346.6-0.2, CTB~37A, G348.5-0.0, and G349.7+0.2.
The infrared colors suggest emission from molecular lines (\nmol\ remnants),
fine-structure lines (\nion), and PAH (\npah), or a combination;
some remnants feature multiple colors in different regions.
None of the remnants are dominated by synchrotron radiation 
at mid-infrared wavelengths.
The IRAC-detected sample emphasizes remnants interacting with
relatively dense gas, for which most of the shock cooling 
occurs through molecular or ionic lines in the mid-infrared.
\end{abstract}


\section{Introduction}

Much of the radiation from supernova remnants is expected to be
emitted in the
infrared range, from heated grains and nebular emission lines. However,
supernova remnants have generally proven to be difficult to detect in the
infrared, especially in the galactic plane where \ion{H}{2} regions are far
brighter. Two attempted infrared supernova remnant
surveys used IRAS observations at 12--100 $\mu$m
and found possible emission from 12 and 14 remnants
\citep{arendt,saken}, respectively, with only 7 in common between the
two surveys, from the sample of  95 remnants in the portion of the
Galactic plane covered in the new survey presented in this paper. 

We present in this paper a new infrared survey of supernova remnants
using the Infrared Array Camera (IRAC) \citep{fazio} on the {\it Spitzer}
Space Telescope \citep{werner}.
In the Galactic Legacy Infrared Mid-Plane Survey Extraordinaire (GLIMPSE)
\citep{churchwell}, the four IRAC arrays---with filters
centered at 3.6, 4.5, 5.8, and 8 $\mu$m and pixels of $1.22''$ 
size---were used to map the inner galaxy within galactic longitudes 
$65^\circ>|l|>10^\circ$ and latitudes $|b|<1^\circ$.
GLIMPSE is a significant advance both because of 
the large increase in angular resolution and sensitivity
as well as covering a new set of infrared wavelengths. 
There are \nsnrinsurvey\ supernova remnants remnants within GLIMPSE
as per the \citet{greencatalog} catalog.
The 5-$\sigma$ sensitivity of GLIMPSE for point sources is 14.0, 12.0, 10.5 and 9.0 mag
at 3.6, 4.5, 5.8, and 8 $\mu$m, respectively.
The raw (1-$\sigma$) surface brightness sensitivity is 0.3, 0.3, 0.7, and 0.6 MJy~sr$^{-1}$
at 3.6, 4.5, 5.8, and 8 $\mu$m, respectively.
At 5.8 and 8 $\mu$m, most of the galactic plane is filled with diffuse emission,
and at 3.6 and 4.5 $\mu$m point-source confusion is significant over much of the 
galactic plane. Thus the primary limitation of this supernova remnant survey is
not instrumental noise but rather confusion from other, overlapping astronomical sources.

\section{Infrared emission from supernova remnants}

To provide a basis for comparison and possible classification of the
infrared colors, we consider here the emission mechanisms expected to dominate
the mid-infrared. Figure~\ref{h2modsconcept}
summarizes the predictions in a color-color diagram.

\begin{figure}
\plotone{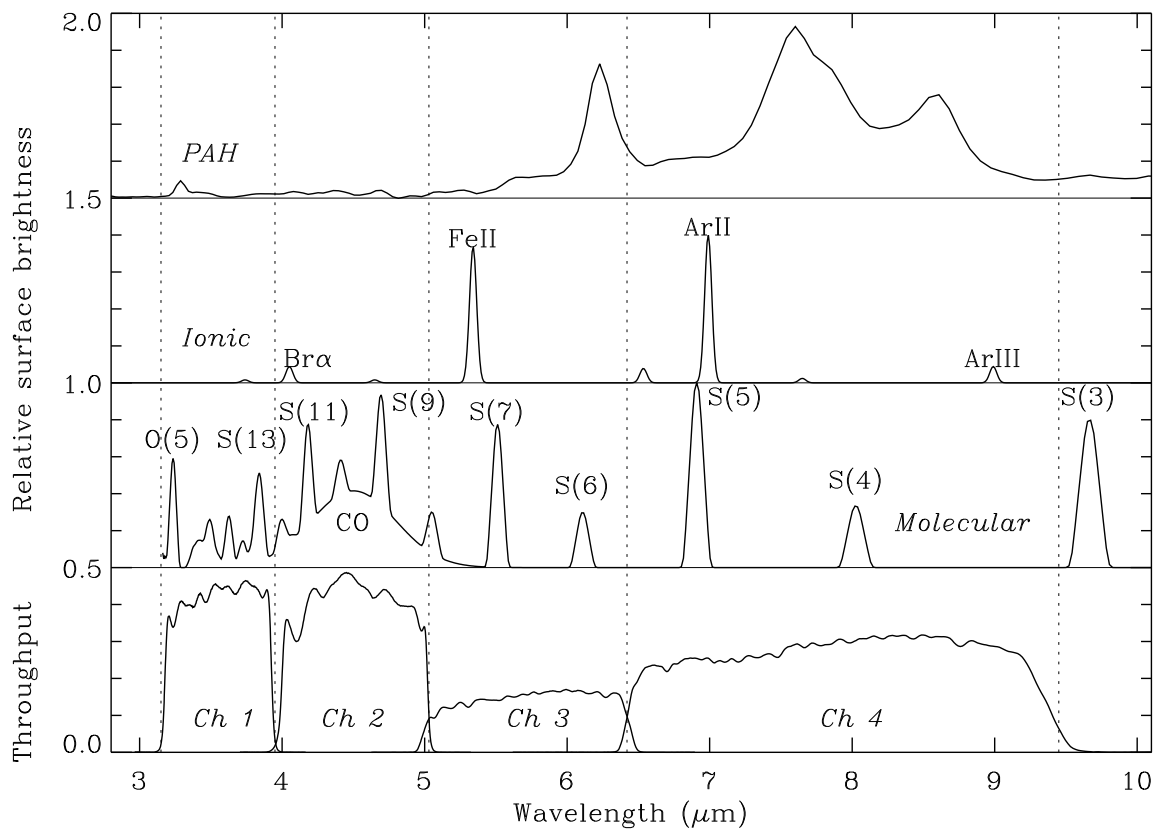}
\epsscale{1}
\figcaption[f1.eps]{
{\it Spitzer}/IRAC spectral response compared to 3 template spectra.
The top curve is the {\it ISO}/SWS spectrum of NGC~7023,
dominated by PAH emission bands;
the second from top is a combination of lines expected from cooling,
ionized gas behind a fast shock; the third from top is a combination
of H$_2$ lines and the CO fundamental band. The lower portion shows
the spectral response for each IRAC channel.
\label{filters}}
\end{figure}

Shocked gas cools through emission lines, and many important emission
lines occur in the mid-infrared. The dominant coolant for shocked
molecular gas, over a wide range of densities, is H$_2$ line emission.
Fast shocks into moderately dense gas
(e.g. 100 km~s$^{-1}$ shocks into gas with density $10^{2-3}$ cm$^{-3}$),
cool via atomic fine-structure lines, 
for which we used a periodic table for fine structure lines
to determine which should be the brightest \citep{rr00}.
The following lines may be significant (in the
indicated channel), {\bf bold} for the brightest:
\ion{H}{2}(8-5)  3.74 $\mu$m (channel 1),
{\bf Br$\alpha$  4.05 $\mu$m (channel 2)},
\ion{H}{2}(8-6)  4.65 $\mu$m (channel 2),
{\bf \ion{Fe}{2} 5.34 $\mu$m (channel 3)},
\ion{Ni}{2} 6.64 $\mu$m (channel 4),
{\bf \ion{Ar}{2} 6.99 $\mu$m (channel 4)},
Pf$\alpha$   7.46 $\mu$m (channel 4), and
\ion{Ar}{3} 8.99 $\mu$m (channel 4).
For the mature supernova remnant 3C~391 with shocks into moderately dense
($n\sim 10^2$ cm$^{-3}$) gas, the 5-15 $\mu$m spectrum showed very bright
\ion{Fe}{2} and \ion{Ar}{2}, in addition to H$_2$ lines \citep{rr02}.
For the very young supernova remnant Cas A,
which is dominated by ejecta and freshly-formed
dust, the 6-16 $\mu$m spectrum showed very bright \ion{Ar}{2}
and \ion{Ar}{3} \citep{arendtcasa}. For the supernova remnant RCW~103, a wide range 
of lines was detected over the wavelengths relevant to IRAC:
H$_2$ lines in all channels, 
Br$\alpha$ in channel 2,
\ion{Fe}{2} in channel 3,
\ion{Ar}{2} and \ion{Ar}{3} in channel 4 \citep{olivaRCW103}.

Except for the youngest ($< 10^3$ yr), 
supernova remnants are dominated by swept-up interstellar matter, 
for which infrared emission 
from dust is inevitable, but the amount and color are not straightforward to
predict. Mid-infrared emission from the interstellar medium
is dominated by polycyclic aromatic hydrocarbon (PAH) bands, 
especially in the IRAC 5.8 and 8 $\mu$m channels. 
Figure ~\ref{filters} shows the spectrum of
the reflection nebula NGC 7023, illustrating how typical interstellar dust
may contribute to the IRAC wavebands.
Grains are sputtered and vaporized in very
strong shocks \citep{jones}, which will reduce the infrared emission
per unit mass. More importantly, in attempting to relate infrared 
features to the supernova remnants, grains are shattered in strong 
shocks \citep{jones}. A size distribution with enhanced smaller grains
will have a higher color temperature, because the smaller grains
are out of thermal equilibrium and emit over a wide range of
higher temperatures \citep{drainlee}. The smallest grains, or
macromolecular PAH may 
be destroyed or altered in even slower shocks and are
largely absent in dense, shocked clumps \citep{rr02}. 
Thus it is nearly impossible to predict the colors of supernova remnants
at IRAC wavelengths: the processes are too complicated (and in
competition with each other), and the shocks span too wide a range
of properties for there to be a typical color. Diffuse
interstellar clouds have a wide range of 12/100 $\mu$m 
ratios---possibly due to shock processing by prior supernova 
blast waves. Given this range of possible initial conditions 
of the grains, it is even more unlikely that the processed grains
will have predictable properties. 
That being said, observations of
infrared emission from SNRs to date have shown little or no evidence
of significant dust emission (PAH or continuum) 
within the wavelengths of the IRAC bands.
We expect shocked dust to 
contribute more in the longest-wavelength IRAC channel than the
others, whether the emission is from macromolecular PAH or from small
grains.

Synchrotron emission can contribute to all wavelengths from the radio to
X-ray, and has been detected in the near-infrared from Cas~A \citep{rhocasa}.
The color of pure synchrotron emission in the IRAC bands would be
approximately \iraccolor 0.6;0.7;0.8;1 . None of the supernova remnants 
we detected have this color. The one with colors closest to
pure synchrotron is for a line of sight for which the mid-infrared
spectrum was measured with {\it ISO} and was dominated by
molecular line emission with negligible continuum \citep{reach391}.
Since many of the SNR in our
sample are very bright in the radio, we expect them to all have at least
faint synchrotron radiation. The expected effect of this synchrotron radiation 
on the mid-infrared colors is to `fill in' the channels that do not have 
significant line or dust emission.

Simple models are presented below for the mid-infrared colors expected
from the three main SNR emission mechanisms described above. A summary
of the derived color/emission mechanism templates is presented in
Figure~\ref{h2modsconcept}.

We now develop simple models to generate template colors for the three
main emission mechanisms. 

\noindent {\it ISM}---First, for the
reflection nebula NGC~7023, the colors in the IRAC channels \iraccolor 1;2;3;4
are \iraccolor 0.054;0.061;0.40;1 . For the \ion{H}{2} region NRAO 530 (as measured
from the Spitzer/IRAC/GLIMPSE data near the supernova remnant 3C~396),
the mid-infrared colors are \iraccolor 0.040;0.046;0.35;1 
(using the same proportion 
notation for channels \iraccolor 1;2;3;4 , wavelengths 
\iraccolor 3.6;4.5;5.8;8 $\mu$m,
which will be used throughout this paper). The origin of these
colors is a combination of PAH and nebular line emission, probably
with a large PAH contribution based on the similarity to NGC~7023.
The \ion{H}{2} region
and reflection nebula colors are very similar and will be difficult 
to distinguish, but for our purpose of classifying unprocessed interstellar
medium this is not important. Figure~\ref{h2modsconcept} shows the 
colors of NGC~7023 and outlines a region of color-color space that
could be attributed to sources with similar spectra.

\noindent {\it Shocked molecules}---The colors of a source dominated by 
molecular emission lines can be estimated using a three-temperature-component
H$_2$ excitation model that matches many lines over a wide range of energy levels
for IC~443 \citep{rho443}. The IRAC colors of a shocked H$_2$ clump
are expected to be \iraccolor 0.42;0.52;0.90;1 . Note the significantly enhanced 
emission in channels 1 and 2. Furthermore, there is a CO fundamental band that
falls within channel 2, with total emission comparable to 
that of H$_2$ in Herbig-Haro shocks with similar
densities and shock velocities. Thus a shocked H$_2$+CO clump would have
colors \iraccolor 0.42;1;0.90;1 . The strong channel 1+2 enhancement turns out to be 
a key to distinguishing supernova remnants interacting
with molecular gas from the 
unrelated interstellar medium or ionized gas.
To allow for variations in the H$_2$ excitation compared to that seen in IC~443,
we computed models for a range of gas temperatures ($>1000$ K) and combinations 
of temperature components that yield at least some of the lines observed toward
IC~443, RCW~103, or HH objects. The infrared emission of two survey remnants
(W~44 and 3C~391) for which we have narrow-band H$_2$ images are
discussed in \S\ref{remnantsec}, validating the usage of IRAC colors
to identify shocked molecular gas.

\noindent {\it Ionized gas}---Pure, ionized hydrogen 
would have IRAC colors \iraccolor 0.25;3.7;0;1, including
the Br and Pf lines listed above, assuming case B recombination at $10^4$ K
\citep{osterbrock}. 
However there will always be some heavier elements in Galactic regions.
We consulted a periodic table for fine structure 
lines \citep{rr00} and used the observed line ratios for RCW~103 
from spectroscopy with the {\it Infrared Space Observatory} \citep{olivaRCW103}
as a guide to the brightness relative to Br$\alpha$. 
Including lines of H, Fe$^+$, Ni$^+$, Ar$^+$, and Ar$^{++}$, 
the predicted colors for ionic shocks
is \iraccolor 0.01;0.10;0.74;1 . Such gas is distinguishable from unshocked ISM and
shocked molecular gas by its
bright channel 3 and very faint channel 1. 
Figure~\ref{h2modsconcept} shows the predicted IRAC colors of RCW~103, together
with a rectangular region that bounds similar regions.
The predicted and observed colors are discussed in the section on RCW~103 below.
Note that if the shocks destroy grains more or less efficiently as in RCW~103, then 
channel 3 will increase or decrease significantly, because 
\ion{Fe}{2} is the dominant ionic contribution to channel 3.

For stellar ejecta, the colors are more difficult to predict and
can have a wide range. 
The infrared colors of young SNRs are expected to vary from
SNR to SNR because the contibution and composition 
of ejecta depend on a number of physical parameters such
as the type of progenitor star, amount of the enriched metal abundances
and the degree of particle acceleration.
For illustration, the ISO spectra of Cas~A
show little or no [\ion{Fe}{2}] emission, which would move
sources vertically in Figure~\ref{h2modsconcept}.
And compared to RCW~103, Cas~A has little \ion{H}{2} emission,
which moves points to the left in Figure~\ref{h2modsconcept}.
From archival ISO data (including TDT 7510064320), the spectrum
of Cas~A shows no bright continuum over the IRAC wavelengths,
and only one bright line ([\ion{Ar}{2}] 6.99 $\mu$m); the
IRAC colors based on the detected spectral lines would be 
\iraccolor 0.01;0.03;0.18;1 ,
with the paucity of channel 1 and 2 emission due to no bright ionic lines
in channel 1 and little expected Br$\alpha$ in channel 2 (due to 
low H abundance). In Figure~\ref{h2modsconcept}, the Cas~A spectrum
falls to the left of the `Ionic' region, but below the `PAH' 
region. None of the supernova remnants we detected in this
survey have the colors of synchrotron emission.

The infrared emission of the survey remnants 3C~391, W~44, 3C~397, and W~49B,
for which we have narrow-band images in near-infrared \ion{Fe}{2} and
H$_2$ filters, is discussed in \S\ref{remnantsec}, 
validating the usage of 
IRAC colors to identify emission from ionic shocks and to distinguish
them from molecular shocks.

\begin{figure}[th]
\plotone{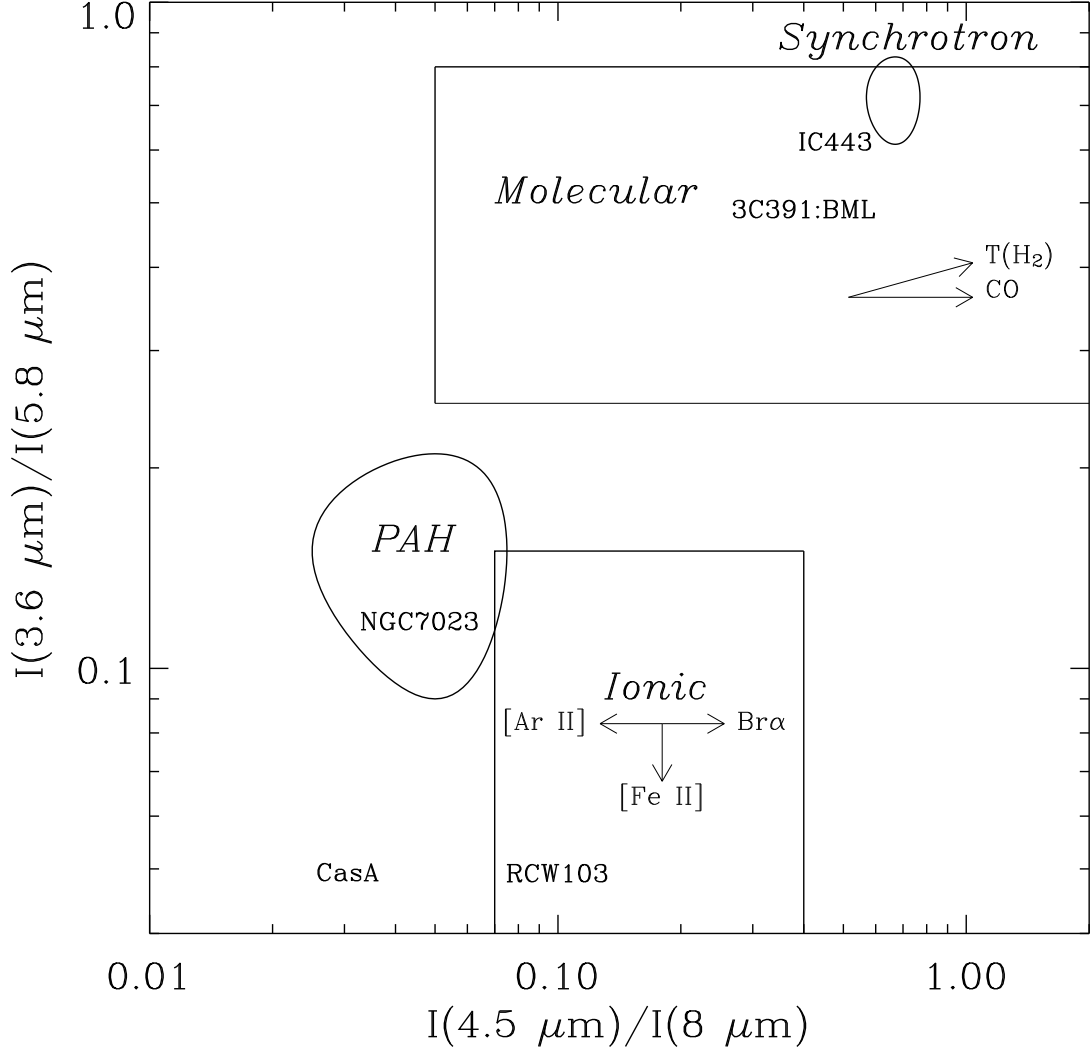}
\epsscale{1}
\figcaption[f2.eps]{Schematic IRAC color-color diagram showing the predicted
colors of the main emission mechanisms expected for supernova remnants
in the mid-infrared: shocked molecular gas, shocked ionized gas, and 
photodissociation regions (PAH). The choice of color axes shown 
here separates the major emission mechanisms best.
The boundaries delineate the
approximate area occupied by a set of models for H$_2$ excitation
and the colors inferred from observed spectra of photodissociation
regions and ionic shocks. 
For molecular shocks two arrows indicate
the color trends for increasing H$_2$ excitation, T(H$_2$), and CO vibrational
emission. For ionic shocks, the arrows show the color trends of increasing Br$\alpha$
4.05 $\mu$m, [\ion{Fe}{2}] 5.35 $\mu$m, and [\ion{Ar}{2}] 6.99 $\mu$m.
The boundaries of the various emission mechanisms are intended to bound the
expected range of physical conditions behind shock fronts. For the molecular
shocks, a range of temperatures from 1000 to 2000 K, and mixtures 
with CO, are included. For ionic shocks, line ratios for RCW~103 were
used as a basis and supplemented with line ratios for fainter lines in M~17.
Material with unusual abundances, such as supernova ejecta or circumstellar material,
can have unusual colors. Some of the template objects described in the 
text (reflection nebula NGC 7023 and supernova remnants Cas A and RCW 103)
are plotted at their approximate IRAC colors.
\label{h2modsconcept}}
\end{figure}

\clearpage

\section{Survey for infrared emission}

\begin{figure}[th]
\plotone{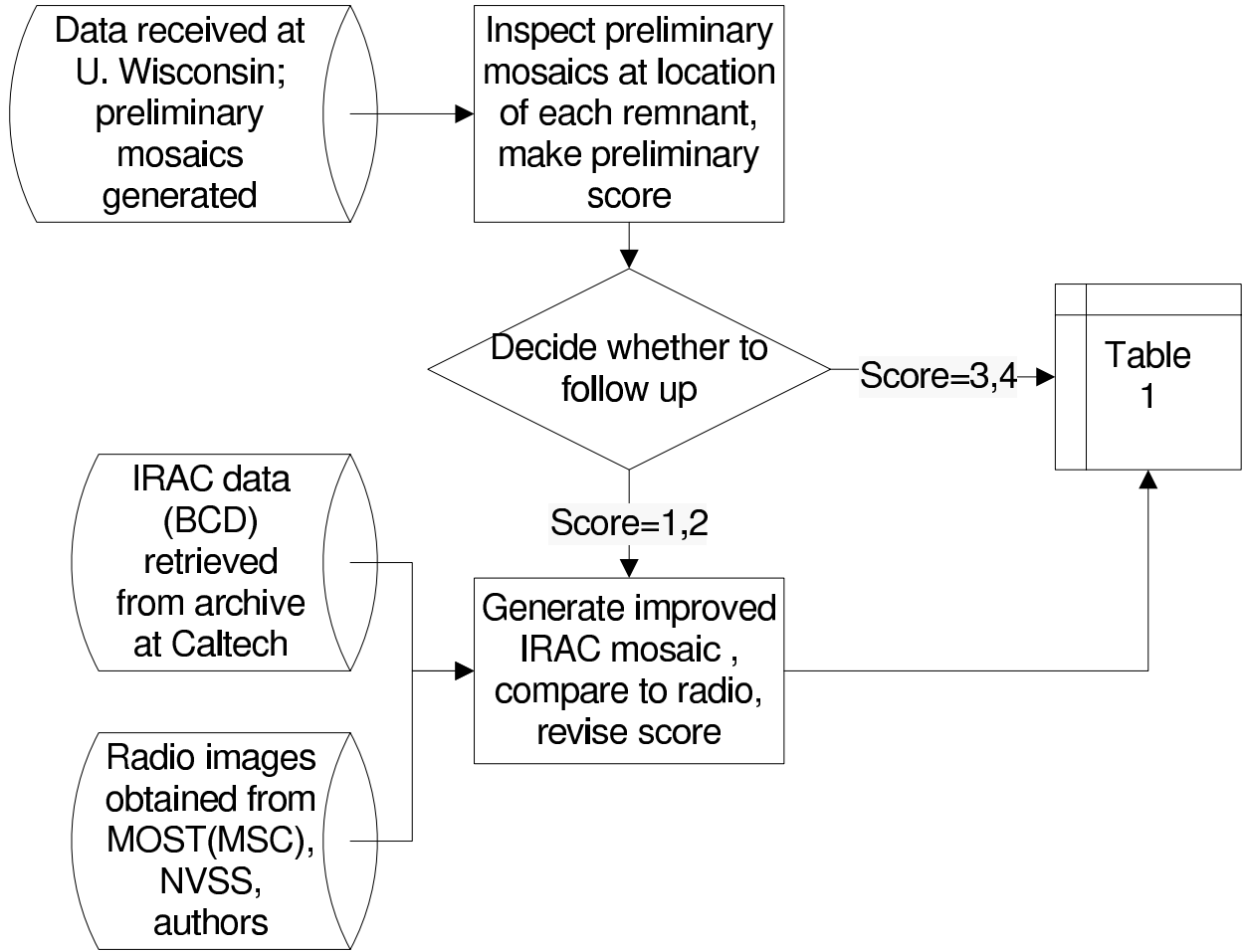}
\epsscale{1}
\figcaption[f3.ps]{
Flowchart of the Spitzer/IRAC/GLIMPSE galactic plane supernova remnant
survey.
\label{flowchart}}
\end{figure}

Table~\ref{surveysummary} presents the list of supernova 
remnants contained within the
boundaries of the infrared survey. The survey procedure is depicted
in Figure~\ref{flowchart}. 
Each remnant was inspected visually on preliminary mosaics generated
immediately upon release of the data by the Spitzer Science Center (SSC).
Greyscale images at 4.5 and 8 $\mu$m were displayed with overlaid
circles depicting the size and location of each remnant in the
\citet{greencatalog} catalog. 
A preliminary score was assigned according to the following scheme:
1=likely detection (with infrared shells or other structures coinciding
with radio, X-ray or optical structure, often with a color distinct from
the diffuse interstellar medium), 
2=possible detection (with some apparent relation between the infrared and
radio image but indistinct colors and too much confusion with unrelated emission), 
3=unlikely detection (but so much confusion with unrelated emission that
there could be significant unrecognized emission from the supernova remnant),
4=not detected.
Remnants with scores of 1 or 2 were selected for follow-up,
and improved mosaics were generated for fields centered on each remnant.
The improved mosaics were generated using the self-calibration 
mosaic technique, allowing for relative background
offsets between images and a flat-field for the array
to be determined simultaneously with the sky mosaic \citep{fixsenarendt}. 
Radio images were obtained from the Molonglo Observatory
Synthesis Telescope (MOST) Supernova Remnant Catalog \citep{MSC},
the National Radio Astronomy Observatory Very Large Array (VLA) 
Sky Survey \citep{NVSS}, individual authors for some
of the individual remnants described in the detailed notes below, or
reprocessed VLA archival data for G11.4-0.1, 3C~391, 3C~396, 3C~397.
The radio contours were superposed on
the infrared images, both the three-color combination of
3.6 (blue), 4.5 (green), and 8 (red) $\mu$m, and a monochromatic
5.8 $\mu$m image. The revised probability of an infrared
counterpart was then determined for each remnant. The score was 1 for
cases where a clear association between the infrared emission and
the radio emission could be made even if the association was not detailed,
for the infrared morphology rarely matches the radio morphology.
The score was 2 for remnants where there is infrared emission that is
suggestive but cannot be convincingly associated with the remnant.
Of the \nsnrinsurvey\ remnants within the survey boundary, \nsnrdetected\ were 
detected (score 1), and \nsnrconfused\ were too confused (score 2). 
In this paper, we concentrate our discussion on remnants with score 1, but 
those with score 2 are worthy of future detailed studies.

\begin{deluxetable}{llrrcllrr}
\tabletypesize{\scriptsize}
\tablecaption{Supernova Remnants in the IRAC/GLIMPSE Survey\label{surveysummary}}
\tablehead{
\colhead{SNR} & \colhead{Name} & \colhead{Size($'$)} &\colhead{Detected?\tablenotemark{a}} &
&
\colhead{SNR} & \colhead{Name} & \colhead{Size($'$)} &\colhead{Detected?\tablenotemark{a}}
}
\startdata
G11.2-0.3  &          & 4 &1 && G308.1-0.7 &		&13 &4  \\
G11.4-0.1  & 	    &8  &3 && G308.8-0.1 &		&25 &2  \\
G12.0-0.1  &	    &7  &3 && G309.2-0.6 &		&14 &3  \\
G13.5+0.2  &	    &5  &3 && G309.8+0.0 &		&22 &3  \\
G15.9+0.2  &	    &6  &3 && G310.6-0.3 &Kes 20B	&8  &2  \\
G16.7+0.1  &	    &4  &3 && G310.8-0.4 &Kes 20A	&12 &1  \\
G18.8+0.3  &Kes 67    &14 &3 && G311.5-0.3 &		&5  &1  \\
G20.0-0.2  &	    &10 &3 && G312.4-0.4 &		&38 &3  \\
G21.5-0.9  &	    &4  &3 && G315.4-0.3 &		&19 &2  \\
G21.8-0.6  &Kes 69    &20 &1 && G315.9-0.0 &		&10 &3  \\
G22.7-0.2  &	    &26 &1 && G316.3-0.0 &MSH 14-57 &20 &3  \\
G23.3-0.3  &W 41	    &27 &2 && G317.3-0.2 &		&11 &3  \\
G23.6+0.3  &	    &10 &3 && G318.2+0.1 &		&37 &3  \\
G24.7+0.6  &	    &21 &3 && G318.9+0.4 &		&20 &3  \\
G24.7-0.6  &	    &15 &4 && G321.9-0.3 &		&27 &3  \\
G27.4+0.0  &4C-04.71  &4  &3 && G322.5-0.1 &		&15 &3  \\
G27.8+0.6  &	    &39 &3 && G323.5+0.1 &		&13 &2  \\
G28.6-0.1  &	    &11 &3 && G327.4+0.4 &Kes 27	&21 &2  \\
G29.6+0.1  &	    &5  &4 && G328.4+0.2 &MSH 15-57 &5  &4  \\
G29.7-0.3  &Kes 75    &3  &3 && G329.7+0.4 &		&36 &2  \\
G31.5-0.6  &	    &18 &3 && G332.0+0.2 &		&12 &4  \\
G31.9+0.0  &3C 391    &6  &1 && G332.4-0.4 &RCW 103	&10 &1  \\
G32.1-0.9  &	    &40 &3 && G332.4+0.1 &Kes 32	&15 &2  \\
G32.8-0.1  &Kes 78    &17 &3 && G335.2+0.1 &		&21 &2  \\
G33.2-0.6  &	    &18 &3 && G336.7+0.5 &		&12 &4  \\
G33.6+0.1  &Kes 79    &10 &2 && G337.0-0.1 &CTB 33	&2  &3  \\
G34.7-0.4  &W 44	    &31 &1 && G337.2-0.7 &		&6  &4  \\
G36.6-0.7  &	    &25 &2 && G337.8-0.1 &Kes 41	&7  &2  \\
G39.2-0.3  &3C396     &7  &1 && G338.1+0.4 &		&15 &4  \\
G40.5-0.5  &	    &22 &4 && G338.3-0.0 &		&8  &3  \\
G41.1-0.3  &3C397     &4  &1 && G338.5+0.1 &		&9  &3  \\
G42.8+0.6  & 	    &24 &4 && G340.4+0.4 &		&8  &4  \\
G43.3-0.2  &W49B	    &4  &1 && G340.6+0.3 &		&6  &2  \\
G45.7-0.4  &	    &22 &2 && G341.2+0.9 &		&19 &4  \\
G46.8-0.3  &	    &15 &3 && G341.9-0.3 &		&7  &4  \\
G49.2-0.7  &W51C	    &30 &3 && G342.0-0.2 &		&10 &3  \\
G54.1+0.3  &	    &2  &3 && G342.1+0.9 &		&9  &4  \\
G54.4-0.3  &          &40 &1 && G343.1-0.7 &		&24 &3  \\
G55.0+0.3  &	    &17 &2 && G344.7-0.1 &		&10 &1  \\
G57.2+0.8  &	    &12 &4 && G345.7-0.2 &		&6  &4  \\
G59.5+0.1  &	    &5  &3 && G346.6-0.2 &		&8  &1  \\
G296.1-0.5 &	    &32 &3 && G347.3-0.5 &		&60 &3  \\
G296.8-0.3 &	    &16 &3 && G348.5+0.1 &CTB 37A	&15 &1  \\
G298.5-0.3 &	    &5  &2 && G348.5-0.0 &		&10 &1  \\
G298.6-0.0 &	    &11 &2 && G348.7+0.3 &CTB 37B	&17 &3  \\
G299.6-0.5 &	    &13 &3 && G349.2-0.1 &		&8  &3  \\
G302.3+0.7 &	    &17 &3 && G349.7+0.2 &		&2  &1  \\
G304.6+0.1 &Kes 17    &8  &1 &&		 &		&   &   \\
\enddata
\tablenotetext{a}{\ Likelihood of mid-infrared counterpart for
the SNR, determined by inspection of the IRAC data and comparison
to existing radio images: 1=detected (18 objects), 2=possibly detected but
confused (17 objects), 3=not detected but confused (44 objects), 
4=not detected at GLIMPSE sensitivity (16 objects).}
\end{deluxetable}  

\clearpage

\subsection{Comparison to previous surveys}

For comparison, 
the {\it IRAS}-based survey by \citet{arendt} detected 12 (17\%) of the remnants
within the portion of the galactic plane included in 
the {\it Spitzer}/IRAC/GLIMPSE survey,
and the independent {\it IRAS}-based survey by \citet{saken} detected 14 (18\%).
Only 7 (10\%) of these remnants were detected in common by the two {\it IRAS}-based surveys:
3C 391, W 49B, G54.1+0.3, Kes 17, G315.4-0.3, G340.6+0.3, and G349.7+0.2.
The new Spitzer/IRAC results presented here are a significant
advance, primarily because of the increase in angular resolution, which
allows better separation of stars, \ion{H}{2} regions, and other interstellar
clouds from the remnants.
The present survey detects 
\nusarendt\ remnants in common with \citet{arendt} and
\nusplussaken\ remnants in common with \citet{saken}. 
There are only \nusarendtsaken\ remnants in common among all 3 infrared surveys:
3C~391, W~49B, Kes~17, G349.7+0.2.

First, let us consider those supernova remnants apparently detected by 
{\it Spitzer} and {\it IRAS}.
We compared the GLIMPSE images to the {\it IRAS} images form the \citet{arendt} catalog.
The {\it IRAS} images of W~44, W~49B, and G349.7+0.2 contain emission plausibly associated 
with the remnant as traced by radio and {\it Spitzer} infrared images.
For 3C~391, the 60 $\mu$m emission shows a believable emission peak from the remnant,
while the other channels are clearly due to unrelated \ion{H}{2} regions. 
For Kes~17 and G11.2-0.3, the structures in the {\it IRAS} images appear to be all due to 
unrelated \ion{H}{2} regions, and their apparent detections are spurious: the dashed
circles that indicate the region within which the fluxes were measured by \citet{arendt}
contain bright \ion{H}{2} regions that are clearly outside the radio shells.

As for the supernova remnants apparently detected by {\it IRAS} and not {\it Spitzer}, 
they are either so confused with \ion{H}{2} regions (e.g. G12.0-0.1) that we cannot tell whether
there are counterparts in either survey, or the {\it IRAS} detections (e.g. CTB~37B) are
likely to be unrelated \ion{H}{2} regions.
In principle, some of the mismatches between {\it IRAS} and not {\it Spitzer}
could be due to the difference in wavelength, since it is possible that
the {\it IRAS} emission is from dust while the {\it Spitzer}/IRAC traces
shocked gas; however, for the specific cases in the present survey.
We attribute the differences between the {\it IRAS} and {\it Spitzer} surveys
to be primarily due to contamination of the {\it IRAS} observations at low
galactic latitudes. Other infrared surveys with higher angular resolution,
for example 24 $\mu$m observations with {\it Spitzer} and future 
far-infrared observations with {\it Hershel} will likely detect many more 
counterparts.

\clearpage

\section{Results for individual remnants\label{remnantsec}}

Table~\ref{detected} lists the \nsnrdetected\ remnants detected by the IRAC/GLIMPSE
survey. Brief descriptions of the mid-infrared emission, and some relevant context,
are given below for each detected remnant, with the subheading 
{\it bold} for detected (score 1) and {\it italics} for possibly-detected (score 2)
remnants.

\begin{deluxetable}{llllllll}
\tablecaption{Properties of detected supernova remnants\label{detected}\tablenotemark{a}}
\tablehead{
\colhead{SNR} & \colhead{Name} & \colhead{Diameter($'$)} & \colhead{region} & \colhead{3.6/8} & \colhead{4.5/8} & \colhead{5.8/8} & \colhead{I8 (MJy/sr)}
}
\startdata
\objectname{G11.2-0.3}  &                       & 4 & SE rim      & 0.34    & 0.48    & 0.58 & 7.5 \\
\objectname{G21.8-0.6}  &\objectname{Kes 69}    &20 & S ridge     & $<0.14$ & 0.91    & 1.4 & 2 \\
\objectname{G22.7-0.2}  &                       &26 & S boundary  & $<0.07$ & $<0.07$ & 0.37 & 23 \\
\objectname{G31.9+0.0}  &\objectname{3C 391}    &6  &BML/OH mas   & 0.18    & 0.36    & 0.67 & 23 \\
	                  &&& NW Fe/radio bar                       & $<0.14$ & 0.41    & 1.7 & 5  \\
\objectname{G34.7-0.4}  &\objectname{W 44}	&31 & E shell     & 0.37    & 0.75    & 0.75 & 11 \\
\objectname{G39.2-0.3}  &\objectname{3C 396}    &7  & W shell     & $<0.10$ & 0.27    & 0.63 & 10 \\
                        &&& central filament                      & $<0.07$ & 0.04    & 0.37 & 27 \\
\objectname{G41.1-0.3}  &\objectname{3C 397}    &4 & N shell      & $<0.07$ & 0.14    & 0.73 & 8\\
\objectname{G43.3-0.2}  &\objectname{W 49B}	&4 & Fe/radio hoop& 0.08    & 0.46    & 1.0 & 8 \\
                       &&& H$_2$ filament                         & 0.16    & 0.42    & 0.76 & 6 \\
\objectname{G54.4-0.3}  &                       &40& N boundary   & 0.04    & 0.04    & 0.33 & 15 \\
\objectname{G304.6+0.1} &\objectname{Kes 17}    &8 & filament     & 0.33    & 0.34    & 0.67 & 5\\
                       &&& shell                                  & 0.12    & 0.30    & 0.39 & 29\\
\objectname{G310.8-0.4} &\objectname{Kes 20A}   &12& SE shell     & 0.18    & 0.18    & 0.61 & 15 \\
\objectname{G311.5-0.3} &	                  &5 & shell        & 0.41    & 0.54    & 0.91 & 9\\
\objectname{G332.4-0.4} &\objectname{RCW 103}   &10& shell        & 0.19    & 0.38    & 0.79 & 15\\
                       &&& filament                               & $<0.07$ & 0.14    & 0.59 & 10\\
\objectname{G344.7-0.1} &	                  &10& shell        & $<0.20$ & 0.23    & 0.82 & 10\\
\objectname{G346.6-0.2} &	    &8 & shell                      & 0.23    & 0.41    & 0.61 & 6\\
\objectname{G348.5+0.1} &\objectname{CTB 37A}   &15& N blob       & 0.11    & 0.32    & 0.52 & 24\\
						             &&& arc          & 0.07    & 0.04    & 0.42 & 62\\
\objectname{G348.5-0.0} &	                  &10& filament     & 0.14    & 0.29    & 0.64 & 14\\
\objectname{G349.7+0.2} &	                  &2 & filament     & 0.05    & 0.14    & 0.59 & 170\\
\enddata
\tablenotetext{a}{The columns 3.6/8, 4.5/8, and 5.8/8 are the ratio of the
surface brightness in IRAC channels 1, 2, and 3 to IRAC channel 4. The emission 
was assumed to be spatially extended, so the colors measured from the images were
multiplied by factors of 1.36, 1.36, and 0.91, respectively \citep{reachcal}.}
\end{deluxetable}  

\begin{figure}[th]
\plotone{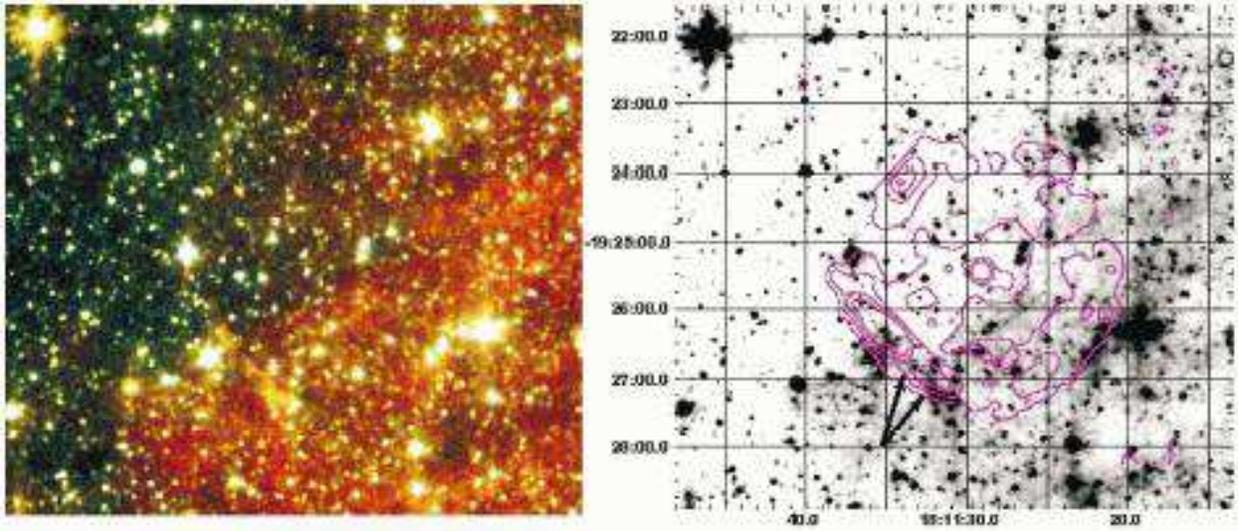}
\figcaption[f4.eps]{
{\it\bf (a)} {\it Spitzer}/IRAC color image of the supernova remnant
G11.2-0.3. The colors are red=8$\mu$m, yellow=5.8$\mu$m,
green=4.5 $\mu$m, blue=3.6 $\mu$m, magenta=5.8$\mu$m - 0.32$\times$8$\mu$m. 
These IRAC images were adaptively smoothed to reduce noise while
preserving angular resolution on bright features.
{\it\bf (b)} {\it Spitzer}/IRAC 5.8 $\mu$m image of the supernova remnant
G11.2-0.3. {\it Chandra} all-energy contours are superposed.
Two infrared filaments are indicated (arrows).
{\it NOTE: this figure was degraded in quality for distribution on astro-ph.}
\label{G11.2combined}}
\end{figure}

{\bf G11.2-0.3}---This compact, circular SNR
has a composite radio morphology, with a clearly-defined, 
steep-spectrum shell (brightest in the SE) combined with a flat-spectrum core.
X-ray observations show a shell (with a thermal spectrum)
similar to that seen in the radio;
a centrally-located pulsar (AX J1811.5-1926) and a pulsar wind nebula
are associated with the SNR \citep{kaspi01}.
Some evidence indicates G11.2-0.3 is associated with the 
supernova of 386 AD,
and the expansion of
the remnant was detected by proper motion of the radio shell \citep{tamroberts};
both of these factors suggest this SNR is relatively young.
The mid-infrared images (Figure~\ref{G11.2combined}a)
reveal a thin filament with three bright segments located within 
the SE rim of the SNR. 
Figure~\ref{G11.2combined}b shows that the filaments
correspond precisely with the two brightest segments of the SE X-ray rim.
The filaments connect to a fainter extension
along most of the eastern radio shell.
More diffuse infrared emission is seen toward the eastern half of the SNR,
though it is unclear if this emission is in fact associated with the SNR. No infrared
emission is seen toward the NW quadrant of the SNR.
There is no infrared emission associated with the pulsar wind nebula.
The filament near \radec 18;11;35.0;-19;26;23 is detected in channels 3 and 4 with
comparable brightness in both channels (10 MJy~sr$^{-1}$), 
but it is not seen in channels 1 or 2.
The blob of emission
at \radec 18;11;31.5;-19;27;16 is detected in all 4 IRAC channels, with color 
ratios \iraccolor 0.3;0.7;1.1;1 . The colors suggest the IRAC emission
from the filaments is dominated by
line emission from shocked gas and certainly not the PAH dust that 
dominates unshocked ISM. Since the remnant is young, 
some of the emission could also arise from ejecta.

\clearpage

\begin{figure}[th]
\plotone{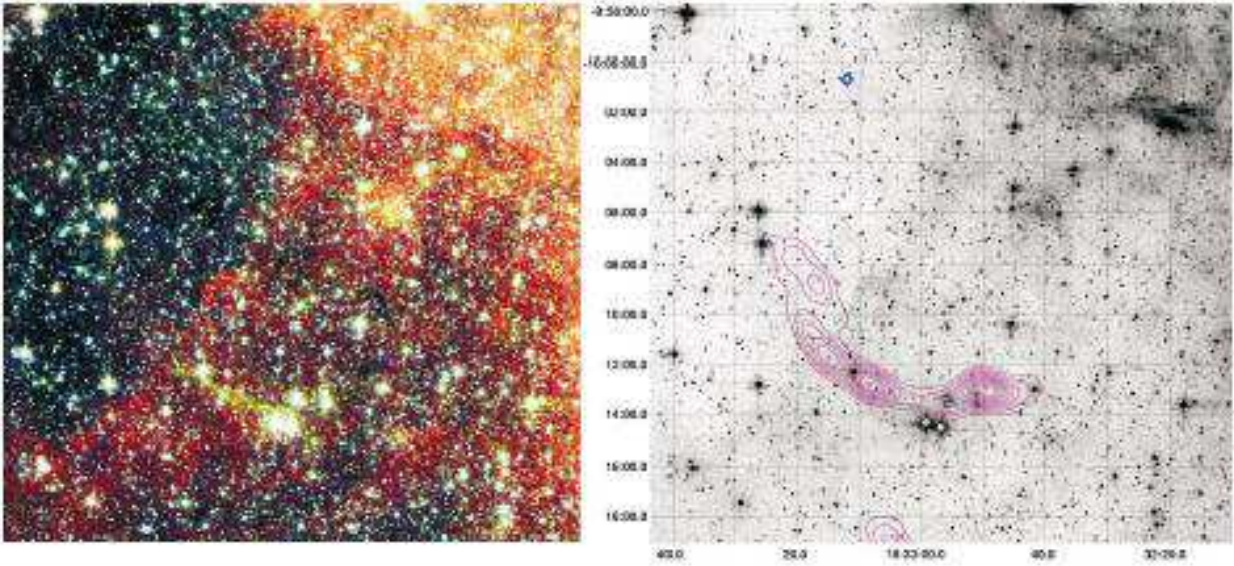}
\figcaption[f5.eps]{
{\it\bf (a)} {\it Spitzer}/IRAC color image of the supernova remnant
Kes 69. The colors are red=8$\mu$m, yellow=5.8 $\mu$m,
green=4.5 $\mu$m, blue=3.6 $\mu$m. 
{\it\bf (b)} {\it Spitzer}/IRAC 5.8 $\mu$m image Kes 69
with radio (NVSS) contours superposed.
The diamond in the NW is the position of an OH 1720 MHz maser.
{\it NOTE: this figure was degraded in quality for distribution on astro-ph.}
\label{Kes69combined}}
\end{figure}

{\bf Kes 69 (G21.8-0.6)}---
Figure~\ref{Kes69combined} shows
a prominent, `green' ridge of mid-infrared emission, 
passing through \radec 18;30;22.0;-10;15;41 in the southern radio shell. 
The IRAC color ratios are \iraccolor ($<0.1$);0.67;1.5;1 ,
which are inconsistent with PAH emission. 
Channels 2 and 3 are most likely dominated by lines from shocked gas.
The exceptionally bright channel 3 emission (relative to channel 4)
could be due to a bright \ion{Fe}{2} line, suggesting
very efficient grain destruction in the shocks. 
The bulk of the radio emission of Kes 69 originates from the 
southeastern shell; likewise, the infrared infrared emission originates
from the same location. The X-ray emission, on the contrary, appears
interior to the radio shell \citep{yusefzadeh03}. 
An OH maser has been detected from the northern part of Kes~69
\citep{greenmaser,yusefzadeh03};
faint, diffuse mid-infrared emission, possibly unrelated
to the SNR, is seen in the vicinity of the OH maser.

\clearpage
\clearpage

\begin{figure}[th]
\plotone{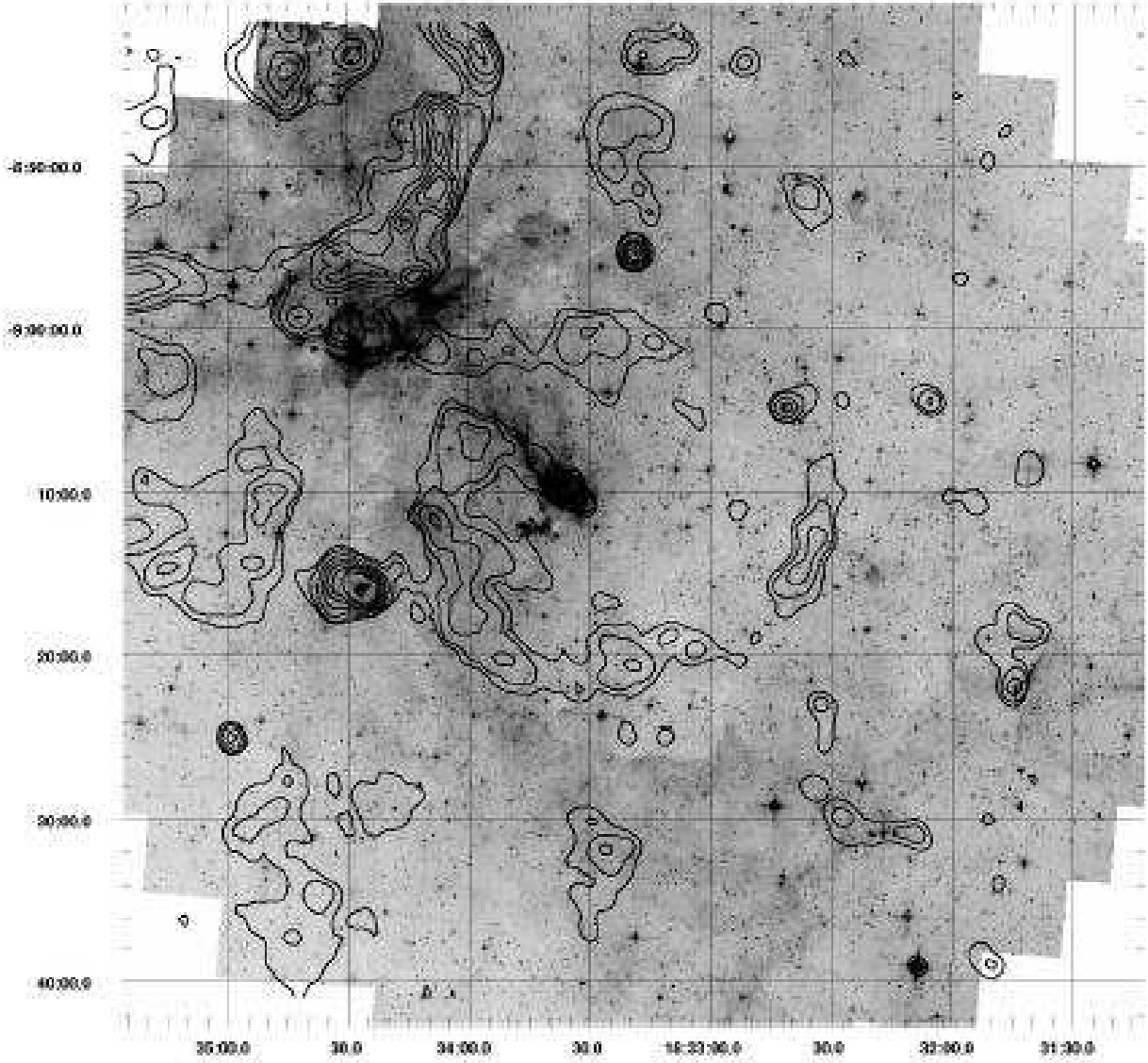}
\epsscale{1}
\figcaption[]{
{\it Spitzer}/IRAC 5.8 $\mu$m image of the supernova remnant
G22.7-0.2, with radio contours from the NVSS.
W~41 is in the upper-left corner of this image.
{\it NOTE: this figure was degraded in quality for distribution on astro-ph.}
\label{G22.7ch3}}
\end{figure}

{\bf G22.7-0.2}---This remnant is located in a field with multiple \ion{H}{2}
regions and is adjacent to W~41. A very bright infrared region 
G22.75-0.25 (\radec 18;33;46.5;-09;10;02 )
is located just E of the center of the remnant. This
region is evident as a very bright compact source
in the 1420 MHz NVSS \citep{NVSS} with flux $\sim 100$ mJy, 
but the source is not evident at 330 MHz \citep{kassim92}, suggesting
it may be thermal emission from an \ion{H}{2} region. Elsewhere along
the supernova remnant shell there is a small probable \ion{H}{2} region
in the NE at G22.73-0.01 (\radec 18;32;42.3;-09;04;43 ). 
A relatively unique region is located in
the W shell at G22.78-0.40, (\radec 18;34;12.7;-09;11;15 ). 
There are numerous mid-infrared
filaments evident at 5.8--8 $\mu$m, and at least one with relatively
bright 3.6--4.5 $\mu$m emission. The infrared emission is probably a
mixture of PAH features and gas lines. The positional coincidence with
the radio shell, filamentary morphology, and infrared colors lead us
to suspect this is a supernova-cloud interaction region, but improved
infrared images or other supporting evidence are needed to properly
classify the nature of this region. 
There is a bright region of infrared emission in between G22.7-0.2 and
the adjacent supernova remnant W~41. 
While this could be a coincidental projection effect,
the location is very suggestive. This could be a case where an interstellar
cloud is being `sandwiched' with impacts from different supernova remnants
(or the progenitors' and cluster members' winds) on either side.
There is a general `deficit' of infrared emission within the radio shell,
clearly evident in Figure~\ref{G22.7ch3}. The region of
decreased infrared emission has a sharp boundary in the southern portion
of the remnant, just south of the NVSS radio contours. The shape of
the southern cavity boundary follows the shape of the remnant too
closely to be a chance alignment. The colors of this southern rim
are very red, possibly due to PAH emission. The morphology and colors
suggest that the observed emission is not from strong shocks,
but instead it may arise from a region evacuated by the progenitor 
(and possibly other cluster members) and currently illuminated by
the remaining cluster members.

\clearpage

{\it W 41 (G23.3-0.3)}---This SNR is located in an incredibly rich field of
infrared emission, with a large \ion{H}{2} region and embedded cluster
(183447.3-083256) in the northern shell, diffuse red emission throughout
the entire region, a
dark lane running roughly  north-south through the entire SNR, 
numerous compact
\ion{H}{2} regions and small shells, and extremely-red sources which
are probably protostars. 
Some of these objects may be related to the SNR progenitor's cluster.  No
detailed association between specific infrared features and the SNR were
noticed (but see discussion of G22.7-0.2 above).

\begin{figure}[th]
\plotone{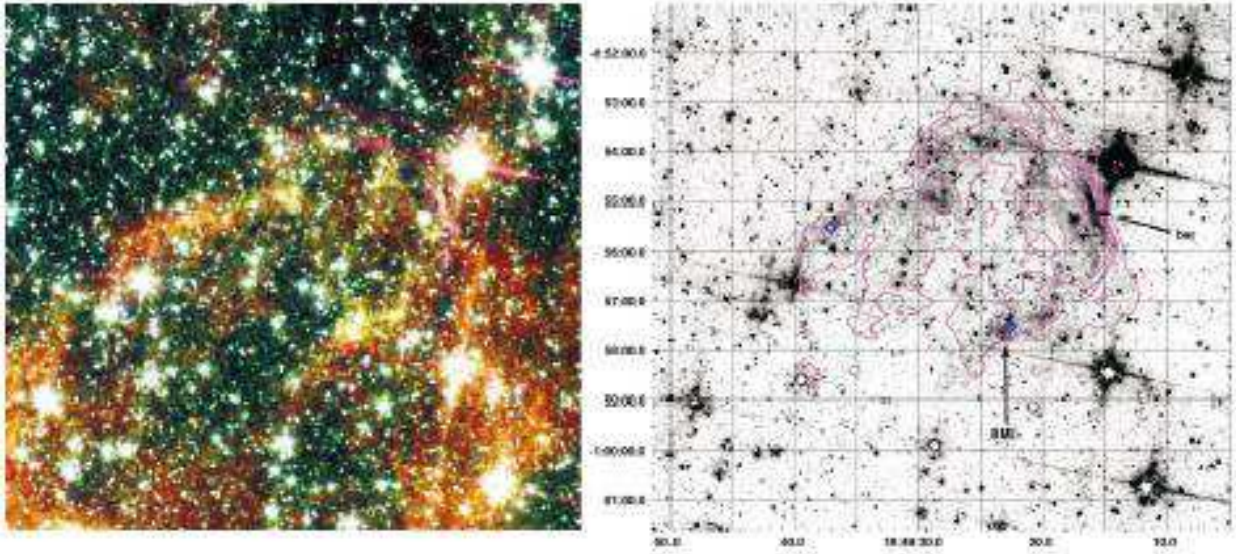}
\figcaption[f7.eps]{
{\it\bf (a)} {\it Spitzer}/IRAC color image of the supernova remnant
3C~391. The colors are red=8$\mu$m, 
green=4.5 $\mu$m, blue=3.6 $\mu$m, 
magenta=$I$(5.8 $\mu$m)$-0.308\times I$(8.0 $\mu$m).
{\it\bf (b)} {\it Spitzer}/IRAC 5.8 $\mu$m image minus a scaled 8 $\mu$m 
image to suppress emission with colors of normal interstellar medium.
The radio contours (magenta) were constructed from VLA 20-cm data at
$16''$ resolution \citep{brogan391}. The two OH masers are indicated
with blue diamonds, and two fetures discussed in the text---the
broad molecular line (BML) region and the radio bar---are labeled.
{\it NOTE: this figure was degraded in quality for distribution on astro-ph.}
\label{3C391combined}}
\end{figure}

{\bf 3C~391 (G31.9+0.0)}---
Figure~\ref{3C391combined} shows the IRAC images
of 3C~391.
This remnant was previously  observed in the mid- and far-infrared using 
ISO: molecular and ionic lines include H$_2$, OH, H$_2$O, CO, [\ion{O}{1}], 
and [\ion{O}{3}] \citep{rr00,rr98} and near-infrared imaging  reveals shocked
H$_2$ and [\ion{Fe}{2}] emission \citep{rr02}. 
In Figure~\ref{3C391combined}a,
two green (channel 2) patches are evident; they are located at the northern
and southern terminus of the bright radio semi-circular radio shell.
The southern one comprises the `broad-molecular-line' (BML)
region 3C~391:BML (\radec 18;49;23.1;-00;57;38 ), where bright near-infrared
H$_2$ emission and broad mm-wave molecular lines were previously
detected \citep{rr02}; it also includes one of the two
1720 MHz OH masers spots associated with 3C391 \citep{frail96}. 
The IRAC color ratios for 3C~391:BML are consistent
with shocked molecular gas.
The northern patch (\radec 18;49;28.8;-00;55;00 ) also has associated near-infrared
H$_2$ emission in a test image taken at Palomar \citep{rr05}; 
reanalysis of the CO and CS spectra reveals no broad molecular lines at 
this position. 
In addition to these patches, fainter and
more extensive emission is evident in channels 2, 3, and 4,
extending around the horseshoe-shaped radio shell. 

The northwestern bar, where the brightest part of the radio
shell is tangent to a giant molecular cloud, is
detected in IRAC channels 2, 3, and 4 and is particularly bright in channel 3.
This bar has very bright near-infrared 1.64 $\mu$m [\ion{Fe}{2}] 
emission \citep{rr02}, and 
IRAC channel 3 contains the 5.34 $\mu$m line of [\ion{Fe}{2}], whose upper 
energy level is the same as the lower energy level of the 1.64 $\mu$m;
the 5.34 $\mu$m transitions should have at least comparable flux 
(and potentially much higher if the gas is cooler than 5000~K).
If we interpret the IRAC channel 3 emission from the bar as entirely
the 5.34 $\mu$m line, its surface brightness is 
$2\times 10^{-3}$ erg~cm$^{-2}$~s$^{-1}$~sr$^{-1}$,
while the 1.644 $\mu$m line brightness at the same position 
is $1\times 10^{-4}$ erg~cm$^{-2}$~s$^{-1}$~sr$^{-1}$; this
line ratio can be reproduced if the shocked gas has $T\sim 2200$~K
(for $n\sim 10^2$ cm$^{-3}$), which is entirely plausible for
the fast J-shocks inferred for this location. Thus for the
bright northwestern bar, IRAC channel 3 is likely to be
[\ion{Fe}{2}] line emission. Some contribution from Ar would be
expected from such shocks in IRAC channel 4:
the ISO 12--18 $\mu$m image was interpreted 
as a sum of [\ion{Ne}{2}] and [\ion{Ne}{3}] lines, and the situation 
should be similar for IRAC channel 4 containing [\ion{Ar}{2}] and [\ion{Ar}{3}].
Comparing the ISOCAM 12--18 $\mu$m and IRAC channel 4 image,
the inferred ratio ([\ion{Ne}{2}]+[\ion{Ne}{3}])/([\ion{Ar}{2}]+]\ion{Ar}{3}])$\sim 2$,
which again is plausibly explained by gas behind a fast J-shock, with
cosmic abundances \citet{rr02}. 

Thus for 3C~391, we find significant
 infrared emission correlated with the radio shell
but with distinct colors: the `green' (enhanced channel 2) regions are
associated with near-infrared emission from H$_2$ and originate from 
shocked molecular gas, while
the `magenta-red' regions (enhanced channel 3) have associated near-infrared
[\ion{Fe}{2}] and orignate from shocked, ionized gas.


\clearpage

\clearpage

{\it Kes 79 (G33.6+0.1)}---Kes 79 was cataloged as a possible
mixed-morphology SNR with centrally-filled X-rays detected
by the ROSAT PSPC surrounded by a
well-defined radio shell \citep{RP98}.  
There is extensive mid-infrared emission near
this SNR, with some features possibly at the distance of the remnant
but no clear evidence for
emission from shock fronts. Interestingly, there is an infrared dark cloud
located near the eastern boundary of the SNR where CO observations suggest
interaction with molecular clouds, although broad molecular lines were
not detected \citep{greendewdney}. There are a number of point sources
within the dark clouds with infrared colors consistent with those
of protostars, and there are also a number of small diffuse structures 
similar to ultra-compact HIII regions. 
No infrared point source coincides with the 
compact central object \citep{seward03}.

\begin{figure}[th]
\plotone{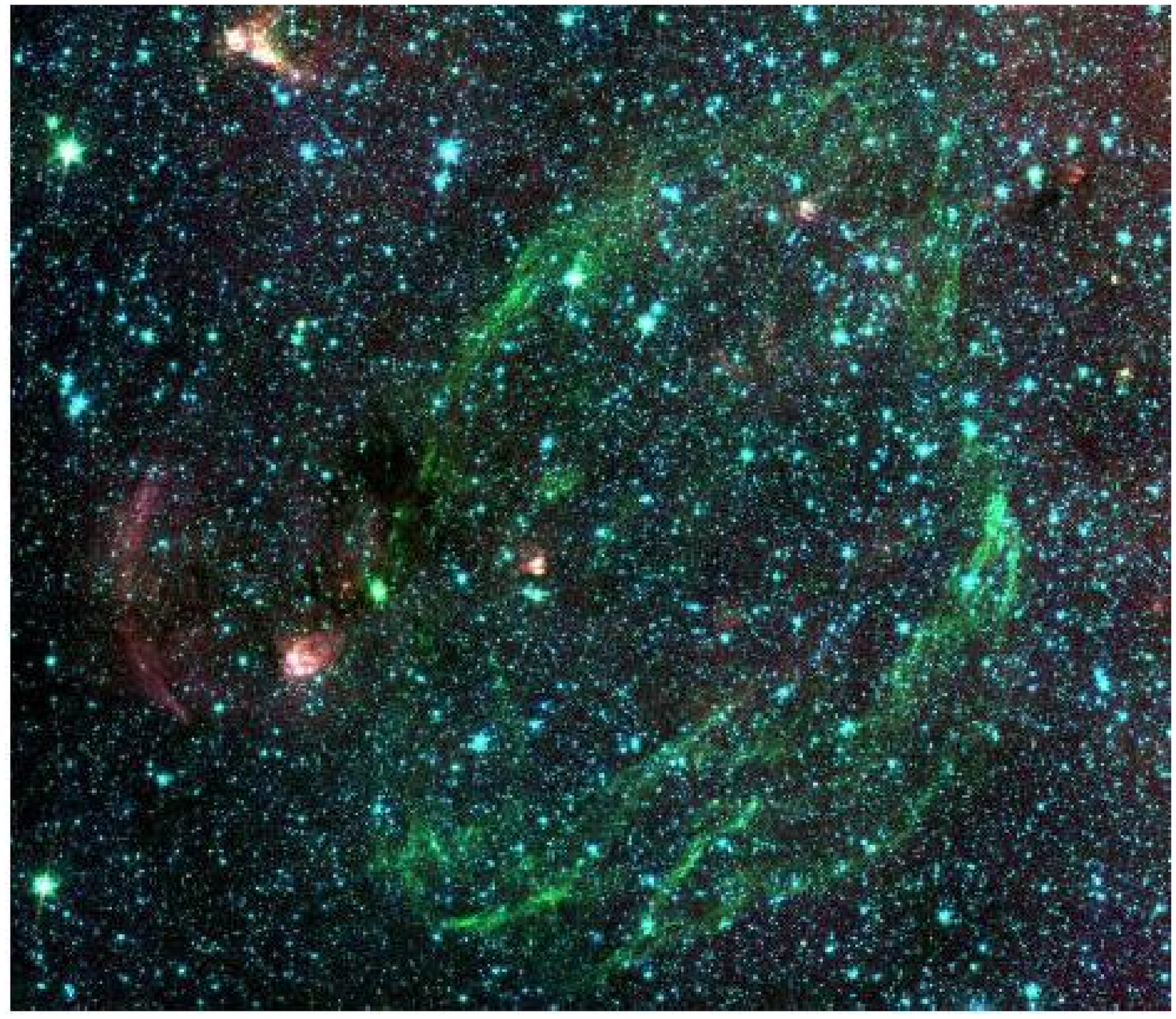}
\epsscale{1}
\figcaption[f8.eps]{
{\it Spitzer}/IRAC color image of W~44. The SNR appears `green' in the IRAC
colors because channel 2 is relatively much brighter in the SNR shell
than in the surrounding ISM (including both the surrounding molecular cloud
and the \ion{H}{2} regions to the E and NE).
{\it NOTE: this figure was degraded in quality for distribution on astro-ph.}
\label{w44color}}
\end{figure}

{\bf W~44 (G34.7-0.4)}---W44 is a mixed-morphology SNR 
featuring centrally-filled, thermal 
X-ray emission surrounded by a well-defined  radio shell \citep{rho44}. The
detections of broad molecular lines and shocked H$_2$ emission 
(Reach et al. 2005 and references therein) 
unambiguously show that the
remnant is interacting with molecular clouds. 
The {\it Spitzer} color image of W44 in Figure \ref{w44color} is one of the best
images from this survey. A large, green, elliptical shell matches the radio
shell rather closely.
In particular, the IRAC channel 2 image emission is almost identical to the
near-infrared H$_2$ images taken for the fields towards the northeastern
and southern portions of the shell \citep{rr05}, indicating that this emission is from
shocked  H$_2$. This is also consistent with the line contribution we
estimated based on the color ratio of the four channel described in
\S2 and validates our predicted IRAC colors for molecular shocks.
There is an infrared-dark cloud at the boundary of the eastern shell
of W44. The patch of red emission south of the dark cloud is likely a
small \ion{H}{2} region based on the ratio of H$\alpha$ and [\ion{S}{2}] \citep{rho44}.
The IRAC channel 3 image is similar to the channel 2 image,
as expected because bright H$_2$ lines in channel 2 will always be accompanied
by bright lines in channel 3. But channel 3 is mixed with the eastern 
\ion{H}{2} region and other ISM emission outside the remnant.
The remnant is barely 
noticeable in IRAC channel 4 due to confusion with unrelated emission.

{\it G36.6-0.7}---There is a possible shell of infrared emission bounding
the SNR toward the North, as well as a dark filament in the N
and clumps in the W and SW. There is no clear evidence for
emission from shock fronts.

\begin{figure}[th]
\plotone{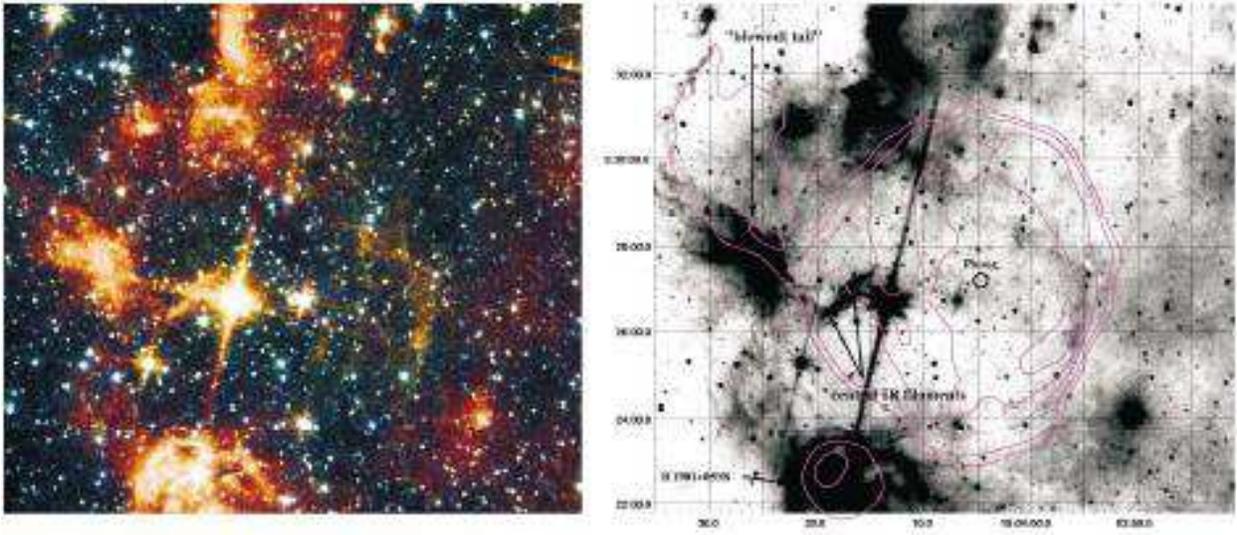}
\figcaption[f9.eps]{
{\it\bf (a)} {\it Spitzer}/IRAC color image of the supernova remnant
3C~396 (G39.2-0.3). The colors are red=8$\mu$m, yellow=5.8 $\mu$m,
green=4.5 $\mu$m, blue=3.6 $\mu$m. 
{\it\bf (b)} {\it Spitzer}/IRAC 8 $\mu$m image of the supernova remnant
3C~396 (G39.2-0.3), with radio contours overlaid. 
The infrared greyscale is logarithmic. The radio
image was generated using 20-cm images from the VLA archive;
contours levels are 1.5, 4.2, 12, 26, 45 mJy/beam (with $15"$ 
resolution). Features discussed in the text are labeled.
{\it NOTE: this figure was degraded in quality for distribution on astro-ph.}
\label{3C396combined}}
\end{figure}

{\bf 3C~396 (G39.2-0.3)}---Infrared emission is detected
from this remnant in three forms. First, faint, filamentary
emission is detected in the western radio shell of this remnant, with 
IRAC colors clearly distinct from normal interstellar emission. 
In Figure~\ref{3C396combined}a, the western shell appears green;
a cut through the western shell 
near \radec 19;03;56.3;+05;25;46 yields IRAC colors \iraccolor ($<0.08$);0.2;0.69;1 ,
suggesting emission from shocked, ionized gas.

Second, there are two very bright infrared filaments,
at \radec 19;04;18.5;+05;20;33 and \radec 19;04;17.0;+05;27;07,
(each $\sim 30''$ long) just inside the eastern radio shell.
The filaments are clearly separated from the 
recently-detected pulsar wind nebula  \citep{olbert3C396}
and are located within a region of exceptionally high radio polarization.
We suspect these bright filaments,
which are highly unusual, are part of the supernova remnant, which is
the only known structure in the interstellar medium at that location.
The color ratios of the filaments are \iraccolor ($<0.05$);0.03;0.4;1 ,
at a location where the 8 $\mu$m surface brightness is 30 MJy~sr$^{-1}$.
These colors are similar to normal interstellar medium, so the filaments
could be photodissociation regions (e.g. compressed filaments that
were shocked long ago, rather than active shock fronts).

Finally, there is faint, diffuse emission surrounding the 
entire radio shell, with some bright, extended regions just 
outside the eastern periphery.
The `blowout tail' discussed by 
\citet{patnaik3C396} from radio data extends eastward from the
remnant and wraps north then back over the top of
the remnant. In the infrared, the `tail' starts with a bright region
at \radec 19;04;26.0;+05;27;55 that is connected to an intricate
set of infrared filaments that follow the radio structure,
with the infrared region displaced somewhat southward.
While this `tail' could be an \ion{H}{2} region,
the remarkably high radio polarization (50\%) 
suggests most of the radio emission
is synchrotron radiation.
Thus the `tail' could be due to energetic
particles in a plume extending through
`hole' in the eastern shell of the remnant.
The bright infrared region near the base of the `tail' 
is detected only in channels 3 and 4, with a ratio 0.3:1 consistent
with photodissociation regions or \ion{H}{2} regions.

\clearpage

\begin{figure}[th]
\plotone{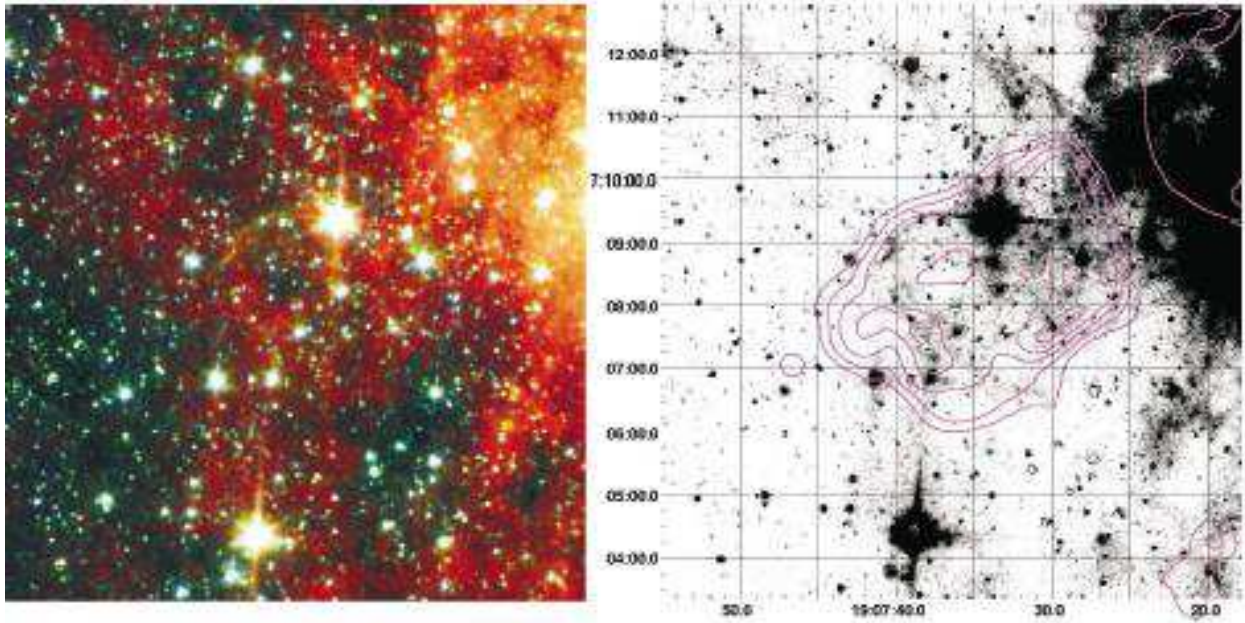}
\epsscale{1}
\figcaption[f10.eps]{
{\it\bf (a)} {\it Spitzer}/IRAC color image
and {\it (b)} 5.8 $\mu$m image of 
3C~397 with radio contours overlaid. 
The infrared greyscale is logarithmic. The radio
image was generated using 20-cm images from the VLA archive;
contours levels are 10, 46, 83, 120 mJy/beam (with $\sim 15"$ 
resolution). 
{\it NOTE: this figure was degraded in quality for distribution on astro-ph.}
\label{3C397ch3}}
\end{figure}

{\bf 3C 397 (G41.1-0.3)}---This small remnant has faint mid-infrared
emission that clearly correlates with portions of the radio shell. 
Figure~\ref{3C397ch3}
shows relatively distinct filaments near
\radec 19;07;40.5;+07;08;48. The infrared
emission was detected in channel 2, 3, and 4, with ratios 
0.1/0.8/1 (for channels 2/3/4)
that are inconsistent with photodissociation regions
and \ion{H}{2} regions but are plausible for fine-structure
lines from shocked gas. The high 5.8 $\mu$m brightness could be
due to efficient grain destruction leading to a bright \ion{Fe}{2}
5.3 $\mu$m line.

\begin{figure}[th]
\plotone{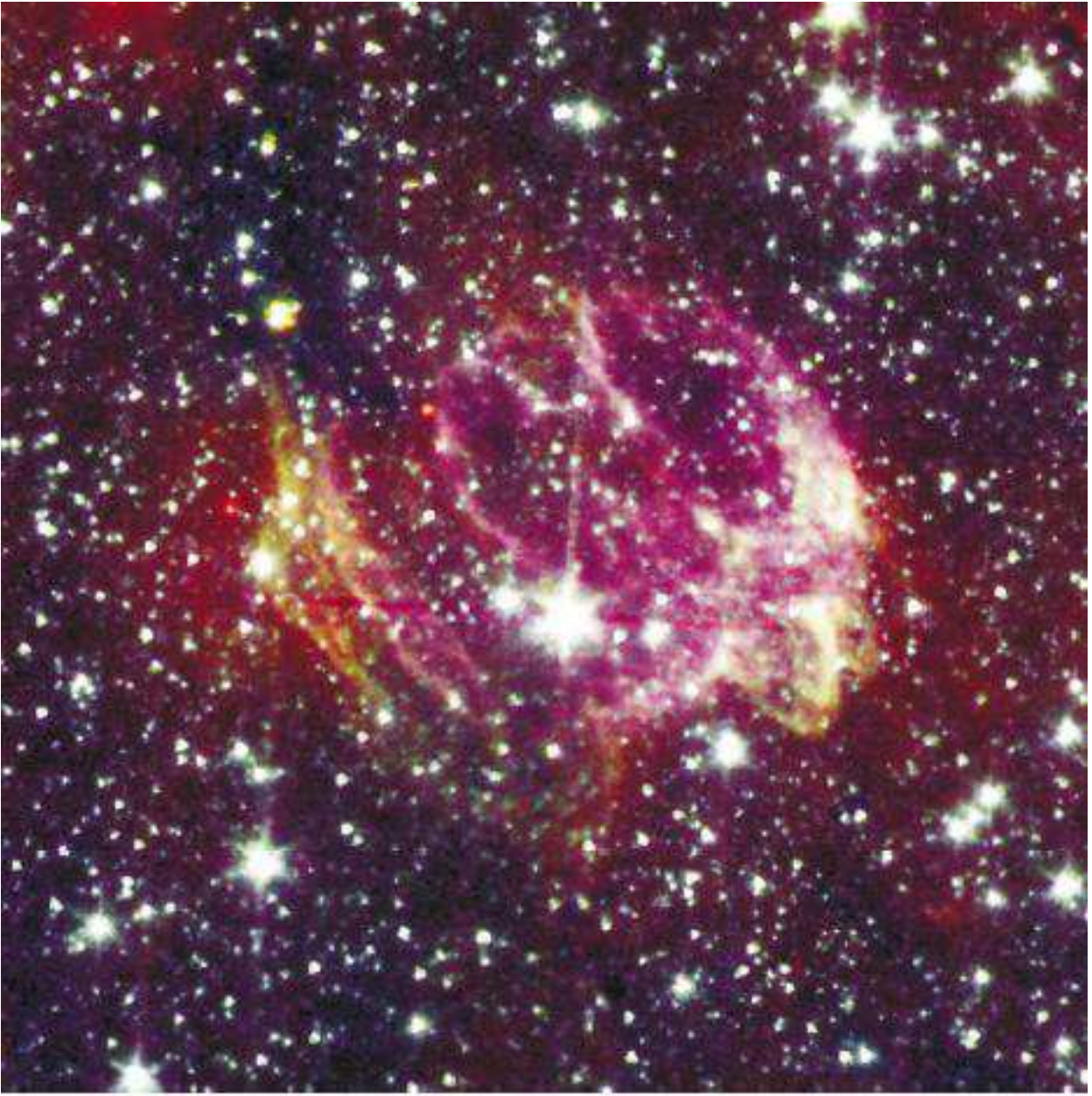}
\epsscale{1}
\figcaption[f11.eps]{
{\it Spitzer}/IRAC color image of the supernova remnant
W~49B, with red=8 $\mu$m, green=4.5 $\mu$m, blue=3.6 $\mu$m,
and magenta=5.8 $\mu$m minus a scaled 8 $\mu$m image. 
There are two distinct types of emitting region, 
the molecular emission is green and is brighter at the
eastern and southwestern extremeties, while the ionic
emission is magenta and is brighter in the radio `loops.'
{\it NOTE: this figure was degraded in quality for distribution on astro-ph.}
\label{W49Bcolor}}
\end{figure}

{\bf W 49B (G43.2-0.2)}---W~49B has a relatively unique structure,
with its radio emission forming a set of curved filaments in 
either a spiral or barrel-hoop morphology \citep{moffettW49B}.
The X-ray emission is thermal with rich line emission, mostly 
attributed to ejecta material, indicating it is a rather young SNR
\citep{hwang00}.
The SNR is clearly detected in the IRAC/GLIMPSE
images. Figure~\ref{W49Bcolor} shows the IRAC color image, with the
emission from the SNR clearly dominating over other nearby diffuse 
emission. There are two very distinct colors 
(green and magenta in Fig.~\ref{W49Bcolor}). 
Much of the SNR has
a filamentary structure, and this part of the IRAC image 
closely follows the radio morphology, with its series of loops.
These appear relatively `magenta' in Figure~\ref{W49Bcolor} due
to very bright emission in channel 3. The color ratios toward
a radio and near-infrared [\ion{Fe}{2}] filament near \radec 19;11;07.0;+09;07;01 
(Table~\ref{detected}) 
suggest line emission from ionic shocks. Another distinct component of
the infrared emission, forming a sort of outer shell toward
the east and southwest, appears relatively `green' in Figure~\ref{W49Bcolor};
the colors toward \radec 19;11;14.3;+09;04;42 (Table~\ref{detected}) are consistent
with lines from shocked molecular gas. The presence of these two 
distinct types of shock was first found in near-infrared, narrow-band
images, which show [\ion{Fe}{2}] 1.66 $\mu$m and radio emission
distributed like the
`magenta' emission in Figure~\ref{W49Bcolor} and H$_2$ 2.12 $\mu$m
emission distributed like the `green' emission in Figure~\ref{W49Bcolor}
\citep{keohane}.
This remnant, together with 3C~391, serve as an empirical
validation of the IRAC color interpretations in this paper
using near-infrared narrow-band imaging.

\clearpage

{\it G45.7-0.4}---There is an infrared filament approximately parallel
to part of the radio shell around \radec 19;16;36.0;+11;08;00, and a potentially
related arc near \radec 19;16;30;+11;14;16, but no clear evidence for emission
from shocked gas or dust.

\begin{figure}
\plotone{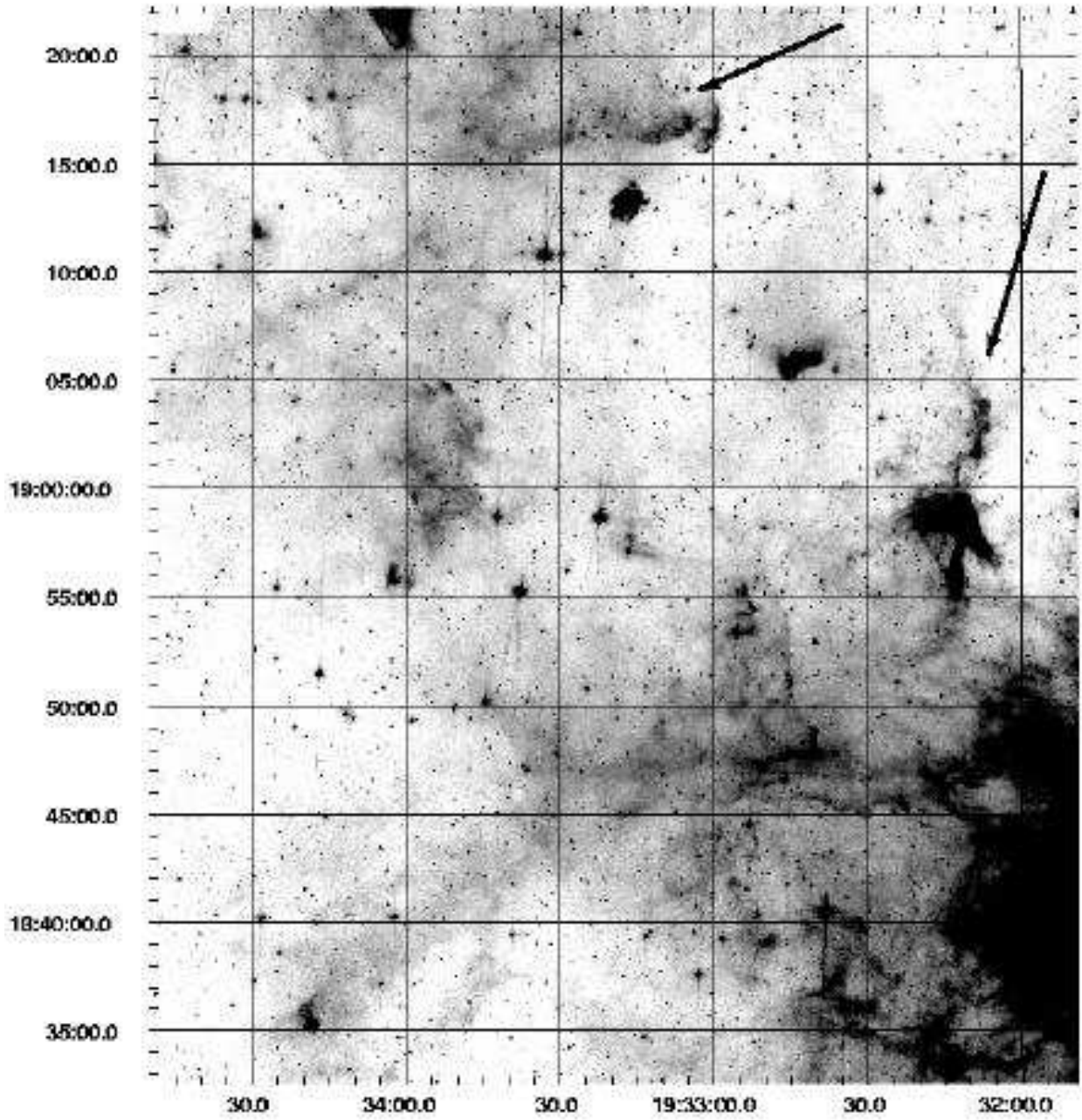}
\figcaption[f12.eps]{
{\it Spitzer}/IRAC 8 $\mu$m image of the supernova remnant
G54.4-0.3. The two filaments discussed in the text are marked with arrows.
{\it NOTE: this figure was degraded in quality for distribution on astro-ph.}
\label{G54.4ch4}}
\end{figure}

{\bf G54.4-0.3}---The cluster containing the progenitor of this SNR 
blew a large stellar wind bubble, into which shocks are now propagating. A shell
of molecular gas was found to be surrounding the SNR, containing protostar 
candidates \citep{junkes}.
Figure~\ref{G54.4ch4} shows two regions where infrared filaments 
follow the radio shell. The features are located along the 
nonthermal, western hemisphere of the radio shell evident at low 
frequencies \citep{velug54}.
These are likely the locations of shocks into 
the wind-blown bubble and/or molecular cloud. The western filament 
is centered on \radec 19;32;07.7;+19;02;56, and the northern filament is centered on 
\radec 19;33;12.6;+19;16;20. The colors of the filaments are similar to photodissociation
regions, so it is not certain whether the emission is due to shocks.
However, there is an infrared shell in both the 5.8 and 8 $\mu$m images,
approximately following much of the radio shell.
Bright \ion{H}{2} regions are located within the shell, with some of
them (e.g. G54.38-0.05) possibly connecting with shell filaments.
These could be a generation of stars that has formed in the shell
surrounding the progenitor of the SNR \citep{junkes}.

\clearpage

{\it G55.0+0.3}---A large diffuse arc ($3'$ wide) possibly relates to the
eastern radio shell, but the association is not clear.

\begin{figure}
\plotone{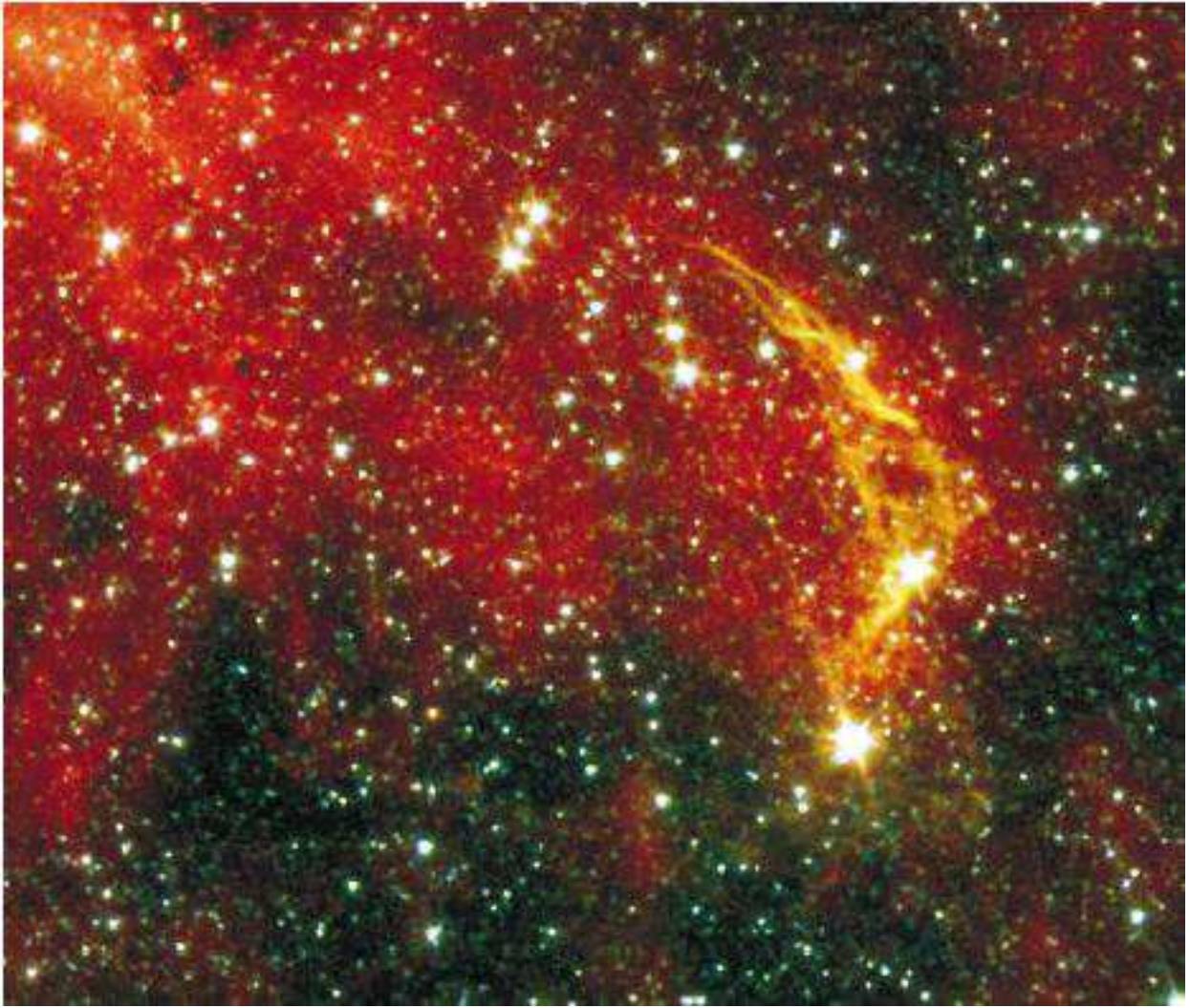}
\epsscale{1}
\figcaption[f13.eps]{
{\it Spitzer}/IRAC color image of the supernova remnant
Kes~17. The image is in equatorial coordinates and has size $13'\times 11'$.
{\it NOTE: this figure was degraded in quality for distribution on astro-ph.}
\label{kes17color}}
\end{figure}

{\bf Kes~17 (G304.6+0.1)}---
Kes 17 is a relatively unstudied SNR. 
The {\it Spitzer}/IRAC images reveal bright mid-infrared emission
shown in in Figure~\ref{kes17color}. The images show bright 
channel 2 emission (green in Fig.~\ref{kes17color} and likely due
to shocked H$_2$) in the  NW rim.
Detailed arc structures are noticeable in the images, and the remnant is
bright in all IRAC channels.  
The IRAC colors of the pair of the thin isolated filament near 
\radec 13;05;46.2;-62;38;33
are somewhat more extreme (brighter in channels 1 and 3, relative to channel 4)
than those toward the brightest part of shell (see Table~\ref{detected}).
Based on the colors and the detailed morphological agreement of the images in
all 4 channels---channel 1 in particular is not expected from
ionic shocks---most of the mid-infrared emission from the shell could
be from molecular shocks.
The radio continuum emission shows clear shells in the
northwest and south. 
Infrared dark clouds and
globules are present north of Kes 17 
(\radec 13;06;33.5;-62;34;00.3) which includes green sources (possible protostars) and
possible compact \ion{H}{2} regions. It is not known whether this star forming
region is associated with (or even at the same distance as) Kes~17.

{\it G308.8-0.1}---This large remnant comprises a very bright northern bar 
(G308.7+0.0) with a southward projection, and a southern arc (G308.9-0.2). 
Interpreted as a single SNR, the diameter is $25'$,
contains the pulsar J1341-6220, and is located at $\sim 7$ kpc, 
behind most of the bright \ion{H}{2} regions in the field \citep{caswell92}. 
The IRAC images contain a wealth of emission; the brightest emission is 
obviously associated with the brightest \ion{H}{2} regions. Two features
are of interest with regard to the SNR. First, following the entire 
extent of the northern bar, there is a dark cloud. 
In fact, the northern bar seems
to nestle within a void of infrared emission. Within the dark cloud, 
there are three embedded sources: \radec 13;40;20.7;-62;16;34 (very red point source,
probable massive protostar), \radec 13;40;57.0;-62;13;05 and 
\radec 13;32;03.0;-62;11;41 (two bright
red blobs probably containing young B stars). It is hard to say whether or
how this dark cloud and its embedded sources are related to the remnant,
particularly if the remnant is relatively distant. The second relevant
infrared feature is extensive filamentary emission within the southern
arc (G308.9-0.2). A remarkable set of bright filaments is near 
\radec 13;42;08.5;-62;32;45;
the IRAC colors of these filaments are \iraccolor 0.02;0.02;0.36;1 , similar to
normal, unshocked interstellar PAH.
The 5.8 $\mu$m image seems to form a shell 
including the southern arc and with a northern boundary far to the south
of G308.7+0.0. Based only on the radio and infrared emission, we might
conclude that there are two separate remnants. However, 
any infrared emission from the northern portion of the SNR is
actually extinguished by the dark cloud that is seen in
projection against G308.7+0.0 
(regardless whether or not the dark cloud and 
G308.7+0.0 are associated).

\def\extra{
\begin{figure}
\figcaption[G308.8color.eps]{
{\it Spitzer}/IRAC color image of the supernova remnant
G308.8-0.1, with the MOST radio contours overlaid. 
G308.7+0.0 is the northern bar, and G308.9-0.2 is the southern
hook-shaped arc. 
The three yellow arrows in the north indicate embedded
sources, and the arrow in the south indicates a remarkable
filament discussed in the text
\label{G308.8color}}
\end{figure}
}

{\it Kes 20B (G310.6-0.3)}---Very dark clouds surround this remnant, including 
remarkable large clouds completely opaque at 8 $\mu$m to the N and E. 
While the present data do not clearly reveal an interaction with these
clouds, an interaction may be occurring to the N where the radio contours
run along the long axis of a dark cloud.
Some 8 $\mu$m emission appears around the border of the remnant but does not have a 
detailed relationship to the radio contours and could be unrelated. 
A notable feature occurs inside the remnant, approximately
along a faint inner radio ridge. It is a thin filament, centered on 
\radec 13;58;08;-62;07;03, with
position angle 67$^\circ$ E of N and length 23$''$,
and it has IRAC colors consistent with a molecular shock.
Without a higher-resolution radio image (to correlate with the infrared filament),
or a more-sensitive infrared image (to find other such features) it is not possible
to tell whether this is a shock front into a molecular cloud or a massive
molecular outflow from a young stellar object. There is no obvious young
stellar object in the IRAC images. Based on the present data we can only
suggest some infrared emission {\it may} be related to the remnant, but
there is not enough evidence to call this remnant a detection.

\begin{figure}[th]
\plotone{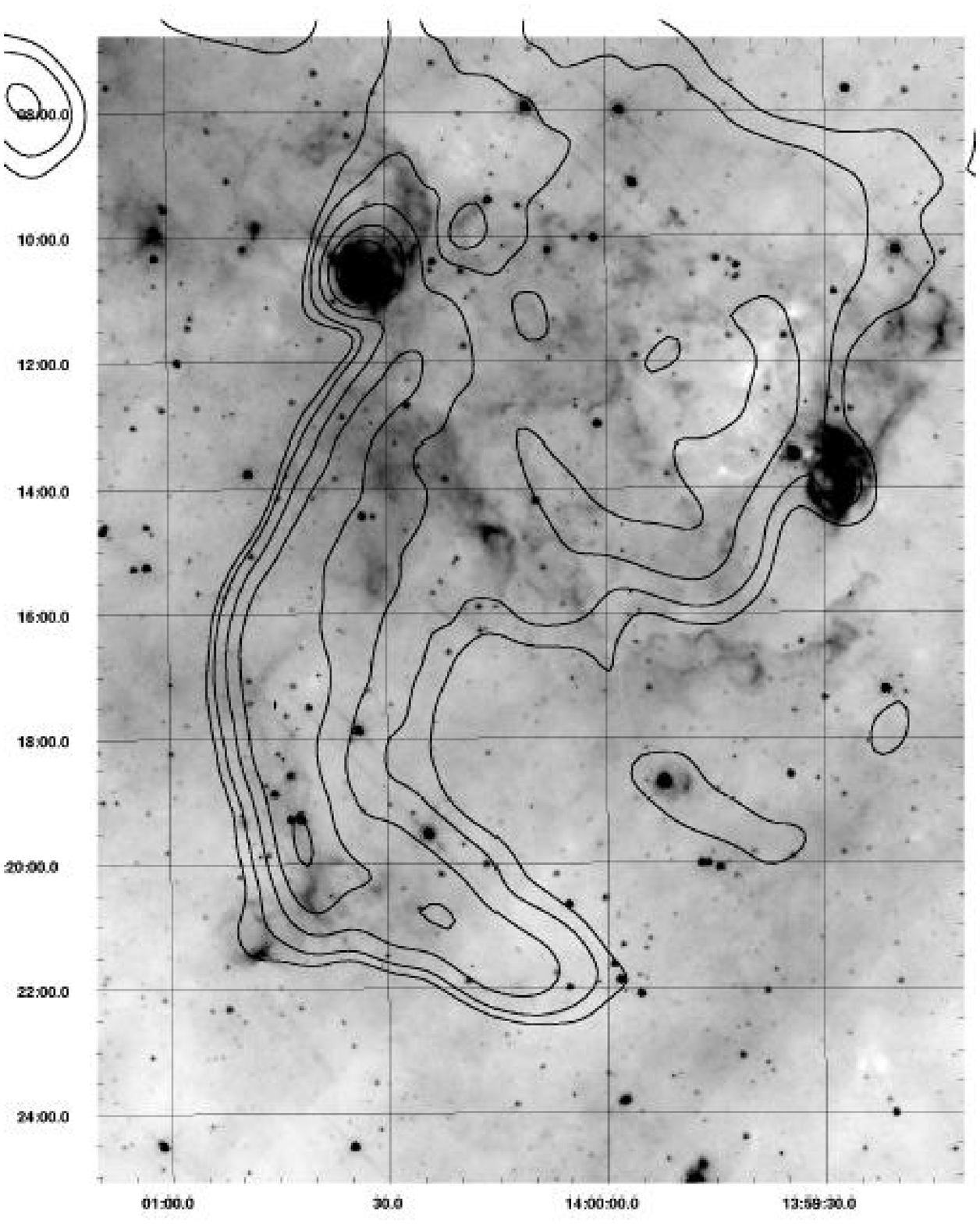}
\epsscale{1}
\figcaption[f14.eps]{
{\it Spitzer}/IRAC image of the supernova remnant
Kes 20A, with radio continuum contours from the MSC overlaid.
{\it NOTE: this figure was degraded in quality for distribution on astro-ph.}
\label{Kes20Abw}}
\end{figure}

{\bf Kes 20A (G310.8-0.4)}---This large remnant 
(Figure~\ref{Kes20Abw}) has a distinctive radio morphology 
defined by a very bright ridge running roughly north-south, defining what could be
the eastern hemisphere of a shell; the western hemisphere is not clear in the
nonthermal radio image. Extensive infrared emission approximately follows the
radio ridge. The infrared emission breaks into many narrow clumps and arcs. 
Based on the infrared image alone, these would appear to be small \ion{H}{2}
regions or reflection nebulae. Indeed numerous similar features that
exist throughout the region are clearly unrelated to Kes~20A. The brightest
such \ion{H}{2} regions are evident in the MOST radio image as individual
sources, clearly distinct from the nonthermal ridge that defines the remnant.
In addition to the emission along the eastern ridge,
similar filamentary infrared emission occupies part of what
would be the interior of the remnant, based on the incomplete-shell radio
morphology. Additional, similar infrared emission is located outside the remnant
and clearly not directly related.
The high density of \ion{H}{2} regions and proximity to the SNR Kes~20B
show this region to be rife with massive stars. Thus the emission that
appears plausibly associated with Kes~20A could be a chance association or
a second-generation association due to younger stars that formed in a
wind-blown bubble generated by the progenitor.
A slice through the SE portion of the
remnant yields infrared colors \iraccolor 0.13;0.13;0.67;1 that do not match the
color templates described above very well but are generally similar to 
those of photodissociation regions, which supports a second-generation
(as opposed to shock-powered) origin for the present infrared emission.

\def\extra{
\begin{figure}[th]
\plotone{../colorfigures/Kes20Acolor.eps}
\epsscale{1}
\figcaption[../colorfigures/Kes20Acolor.eps]{
{\bf\it NOTE: THIS FIGURE IS NOT REFERENCED AND WILL BE DELETED}
{\it Spitzer}/IRAC image of the supernova remnant
Kes 20A, with red=8 $\mu$m, green=4.5 $\mu$m, blue=3.6 $\mu$m.
\label{Kes20Acolor}}
\end{figure}
}

\clearpage

\begin{figure}[th]
\plotone{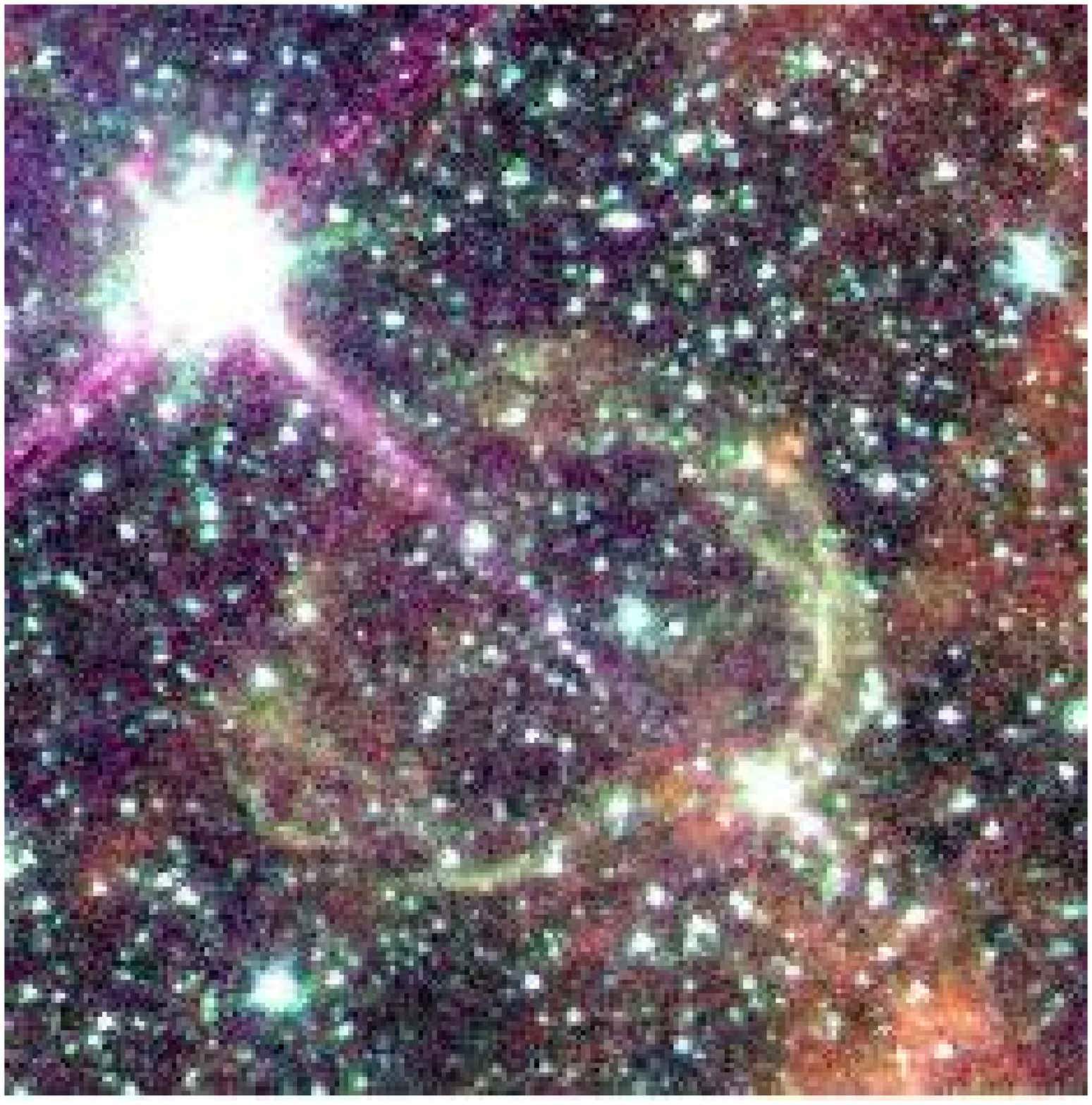}
\epsscale{1}
\figcaption[f15.eps]{
{\it Spitzer}/IRAC color image of the supernova remnant
G311.5-0.3, with red=8 $\mu$m, green=4.5 $\mu$m, blue=3.6 $\mu$m,
and magenta=5.8 $\mu$m minus a scaled 8 $\mu$m image. 
The entire shell of the SNR is evident, with a double-stranded
morphology.
{\it NOTE: this figure was degraded in quality for distribution on astro-ph.}
\label{G311color}}
\end{figure}

{\bf G311.5-0.3}---This shell-type remnant is one of the most easily
detected in the present survey, because it so clearly follows the 
entire radio shell and it is in a relatively uncluttered region.
Cohen and Green (2001) compared MOST observations with
8.3 $\mu$m observations made with the Midcourse Space Experiment (MSX)
and reported no detection. Figure~\ref{G311color} shows the
Spitzer/IRAC image, which is significantly
deeper than MSX, so the SNR is evident in all 4 channels.
The IRAC images show a nearly complete shell of infrared emission corresponding 
well with the radio shell. 
Bright narrow filaments line the western edge of the SNR, with two bright
ridges giving a `braided' appearance
over a significant portion of the western shell.
The infrared shell is particular bright near \radec 14;05;21.9;-61;58;06, and
a less-confused filament nearby (\radec 14;05;22.6;-61;57;22) has 
IRAC color ratios \iraccolor 0.3;0.4;1;1 suggesting the IRAC emission is
from shocked gas.
Two bright, red, compact sources at \radec 14;05;24.3;-61;57;07 and 
\radec 14;05;23.5;-61;56;58
are possibly Class 0 protostars located just at the edge of the SNR.

{\bf Kes 27 (G327.4+0.4)}--- The mid-infrared image shows a set of bright filaments
with diffuse emission centered on a radio peak within the remnant near J154912-534420
(G327.39+0.47). 
Figure~\ref{kes27ch4bwlab} shows the correspondence between
this region and the radio shell.
This filamentary source
could be the location of shocks into a dense cloud, possibly containing
an embedded massive star, so we consider it in more detail. The
MOST 843 MHz image \citep{MSC} shows a resolved but simple (centrally
condensed) peak with a FWHM of 2.2$'$ and a flux of approximately 0.4 Jy
(from aperture photometry centered on the source with annular 
background removal).
The Parkes 5 GHz image \citep{milnedickel75} 
does not show a corresponding structure,
at least partially due to its low angular resolution (4.4$'$). If the
source is an \ion{H}{2} region with thermal spectrum, the flux at
5 GHz would be $\sim 0.2$ Jy and the brightness temperature
$\sim 0.3$ K after diluting to the Parkes 5 GHz beam. On the other
hand, if the source is nonthermal with spectral index 0.7, then 
the brightness temperature at 5 GHz would be 0.06 K. 
There is no structure in the 5 GHz image above 0.2 K, which is
consistent with nonthermal emission and marginally inconsistent
with thermal emission.
\citet{mccluregriffiths} show a 1420 MHz continuum 
image of the remnant (their Fig. 13). The radio peak corresponding
to the infrared region is evident in their image, with a flux of
approximately 0.2 Jy (counting contours). The 1420 MHz and 843 MHz
contour maps are nearly identical after a linear transformation,
indicating that the infrared filamentary source has about the
same spectral index as the rest of the remnant emission.
Improved radio continuum observations
are needed to assess the nature of the source.
The mid-infrared colors of this region are more typical of 
PAH-dominated photo-dissociation regions than shocked gas.
\citet{mccluregriffiths} show from 21-cm line observations
that there is an \ion{H}{1} ridge just outside the southwestern
radio contours. There are no obvious protostar candidates in the region,
so the impact has evidently not triggered star formation; this is
not surprising given the remnant is thought to be relatively
young, with estimates of 2400 yr \citep{mccluregriffiths} 
and 3500 yr \citep{sewardkes27} [cf. $>8\times 10^4$ yr \citep{enoguchi}].
The true nature of this region awaits future investigation.

\def\extra{
\begin{figure}[th]
\plotone{../colorfigures/Kes27color.eps}
\epsscale{1}
\figcaption[../colorfigures/Kes27color.eps]{
{\it\bf NOTE: THIS FIGURE IS NOT REFERENCED AND MAY BE DELETED}
{\it Spitzer}/IRAC color image of the supernova remnant
Kes 27. The colors are red=8$\mu$m, 
green=4.5 $\mu$m, yellow=5.8 $\mu$m, blue=3.6 $\mu$m,
magenta=5.8$\mu$m-$0.3\times$8$\mu$m.
\label{Kes27color}}
\end{figure}
}

\begin{figure}[th]
\plotone{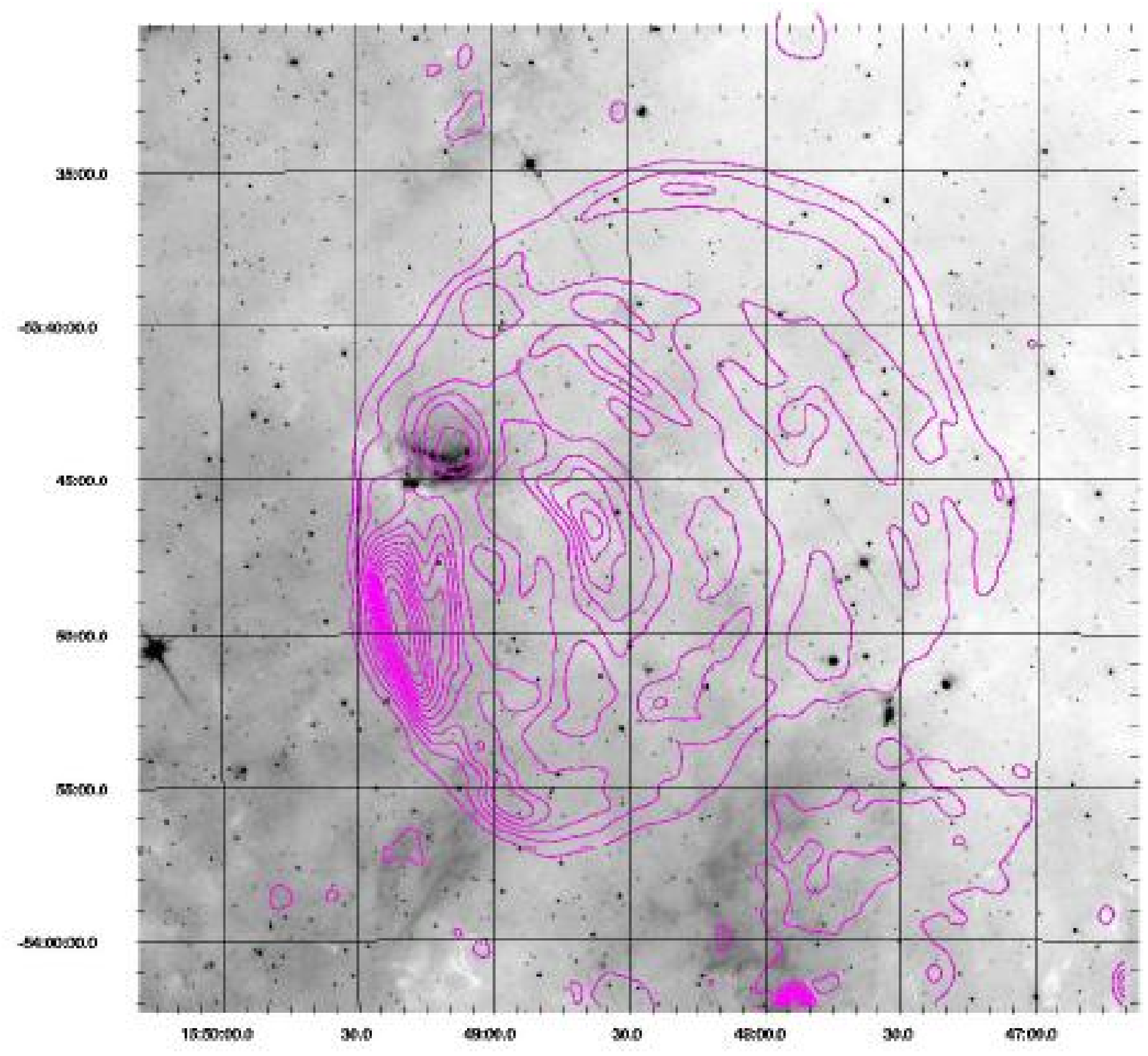}
\epsscale{1}
\figcaption[f16.eps]{
{\it Spitzer}/IRAC 8 $\mu$m image of the supernova remnant
Kes 27, with radio contours from the NVSS overlaid.
{\it NOTE: this figure was degraded in quality for distribution on astro-ph.}
\label{kes27ch4bwlab}}
\end{figure}

\clearpage

{\it G329.7+0.4}---The SW corner of this remnant has bright extended
mid-infrared emission that follows some of the radio structures, in particular 
the SW corner of the remnant and a bright spur penetrating into the 
remnant from \radec 16;00;39.9;-52;30;46 to \radec 16;01;10.3;-52;25;24 
(and beyond). 
The corner and spur have similar brightness to some \ion{H}{2} regions,
but their orientation along a radio filament that appears to be part of
the remnant makes them plausibly part of the remnant. On the other hand
there are some mid-infrared spurs that appear related to this
same region but directed well outside the remnant.
The colors of the bright bar are \iraccolor 0.03;0.02;0.33;1
(toward a location where the 8 $\mu$m surface brightness is 
66 MJy~sr$^{-1}$) look like PAH emission or an \ion{H}{2} region.
A `shell' of mid-infrared shell surrounds the remnant on the eastern
and southern sides, connecting to the bright SW corner and the 
western side of the remnant. This `shell' breaks into some moderately
bright patches that may be individual \ion{H}{2} regions or 
photodissociation regions from B stars. The overall box-like shape
of the region is very similar to the outer boundary of the nonthermal
emission from the supernova remnant. Therefore, we suspect that it
is in fact related to the remnant, even though it is unlikely to be
emission from shocked dust and gas. Instead it probably represents
a fossil shell, due to the stellar winds and supernovae of the
previous generation of massive stars, into which the present
remnant is now expanding. A relatively diffuse part of the shell
has colors \iraccolor 0.05;0.04;0.6;1 (toward a location where the 8 $\mu$m 
surface brightness is 11 MJy~sr$^{-1}$), most similar to PAH 
regions or \ion{H}{2} regions.

\begin{figure}[th]
\plotone{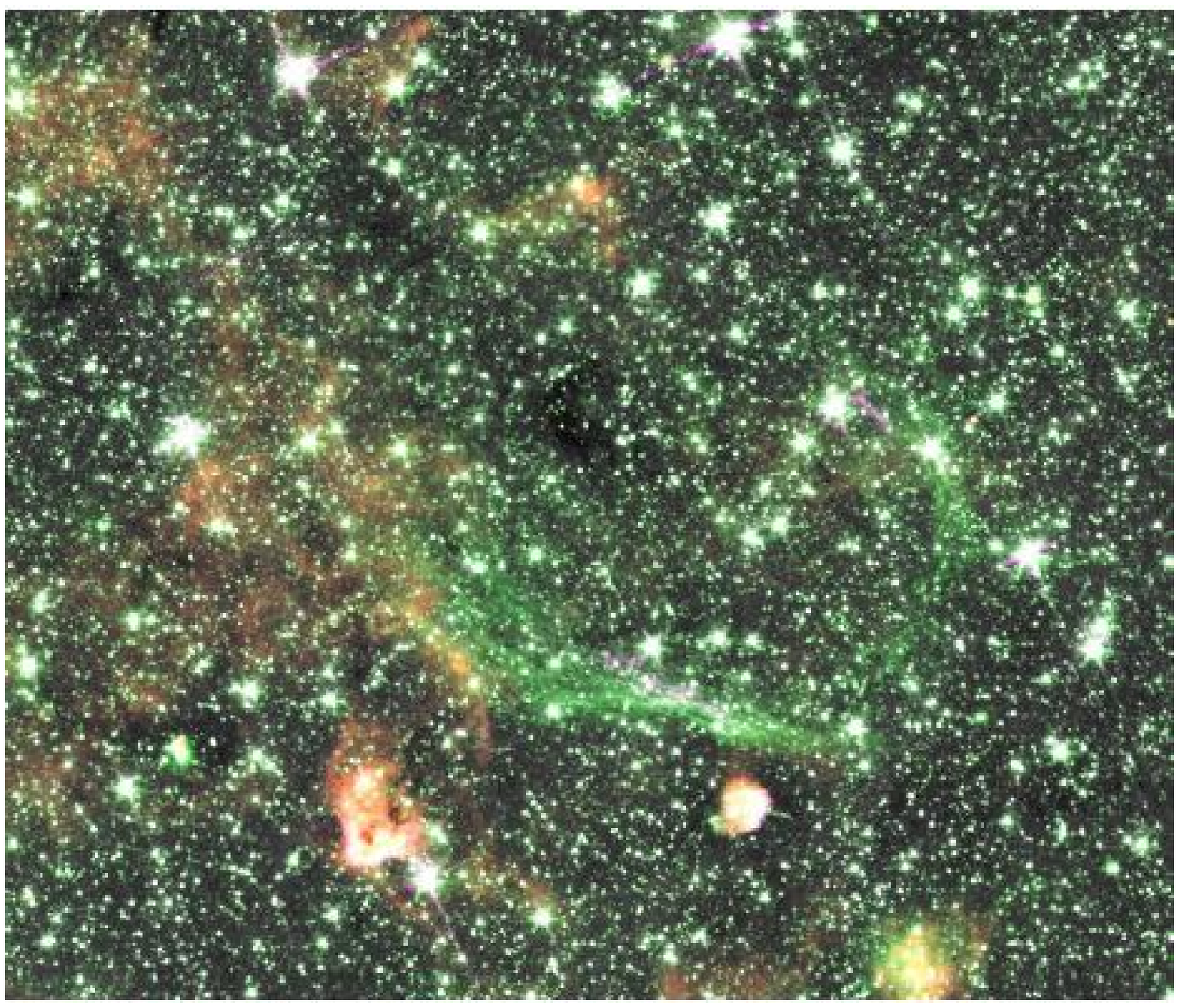}
\epsscale{1}
\figcaption[f17.eps]{
{\it Spitzer}/IRAC color image of the supernova remnant
RCW 103, with red=8 $\mu$m, green=4.5 $\mu$m, blue=3.6 $\mu$m,
and magenta=5.8 $\mu$m minus a scaled 8 $\mu$m image. The SNR
is most clearly separated from the diffuse galactic emission 
by its 4.5 $\mu$m emission, but it is detected in all
4 IRAC channels. There are at least two distinct emitting regions;
the dominant one (seen in all 4 channels) is green in this 
rendition, while a fainter one (most noticeable at 5.8 $\mu$m)
is magenta in this rendition.
{\it NOTE: this figure was degraded in quality for distribution on astro-ph.}
\label{RCW103color}}
\end{figure}

{\bf RCW 103 (G332.4-0.4)}---
The IRAC images (Figure~\ref{RCW103color}) in all 4 channels show relatively bright, diffuse emission
associated with the SNR RCW 103. In particular channel 2 shows 
strong emission at the southern shell and faint emission in the
northwest and western shell. H$_2$, [\ion{Fe}{2}] and [\ion{Ar}{2}] lines were
previously detected in ISO spectra, and the emission was spatially resolved
in the ISOCAM FOV 1\farcm 5 image  \citep{olivaRCW103}.
Therefore, we measure the IRAC color from two positions
of the SNR. 
A slice through the filament near \radec 16;17;32.8;-51;06;28 yields colors
\iraccolor 0.14;0.28;0.86;1 (with 8 $\mu$m brightness 14.5 MJy~sr$^{-1}$),
most likely dominated by shocked molecules (but with some ionic
contribution in channel 3). Much of the southern rim has such colors.
A thin filament near \radec 16;17;45.7;-51;04;58 (Table~\ref{detected})
is undetected in channel 1 despite being very distinct in channel 3
indicating  ionic shocks. The two types of emitting regions can be discerned
in Figure~\ref{RCW103color}, with the primarily-molecular shocks 
appearing green and the primarily-ionic shocks appearing magenta. 
There are a number of \ion{H}{2} regions and dark clouds
surrounding RCW~103.

\clearpage

{\it Kes~32 (G332.4+0.1)}---There is no mid-infrared emission corresponding
to the radio contours from MOST. However, there is a very bright
infrared \ion{H}{2} region in the southwestern part of the remnant.
The \ion{H}{2} region is centered near \radec 16;15;37.9;-50;44;06 and has
a central cavity containing what appears to be a cluster of 
young stars. Protrusions extend from this \ion{H}{2} region in 
many directions, including narrow filaments and what appears to be
a very large `plume' extending to the northwest of the \ion{H}{2}
region, directly across the western half of Kes~32. It is not clear
which features are associated with the remnant and which are associated
with the \ion{H}{2} region. However, the high-resolution Spitzer/IRAC 
image shows the `plume' connects directly to the \ion{H}{2} region and
is most likely associated with it. This contradicts an 
earlier conjecture that the plume, which can been seen in radio
images, could be a jet of energetic particles from the stellar
remnant of Kes~32 \citep{rogerKes32}. Instead, we suspect that the
plume originates from the \ion{H}{2} region and is being 
ionized by the current generation of young stars within it.

\def\extra{
{\it G335.2+0.1}---

{\it Kes 41 (G337.8-0.1)}---

{\it G340.6+0.3}---
}

\begin{figure}[th]
\plotone{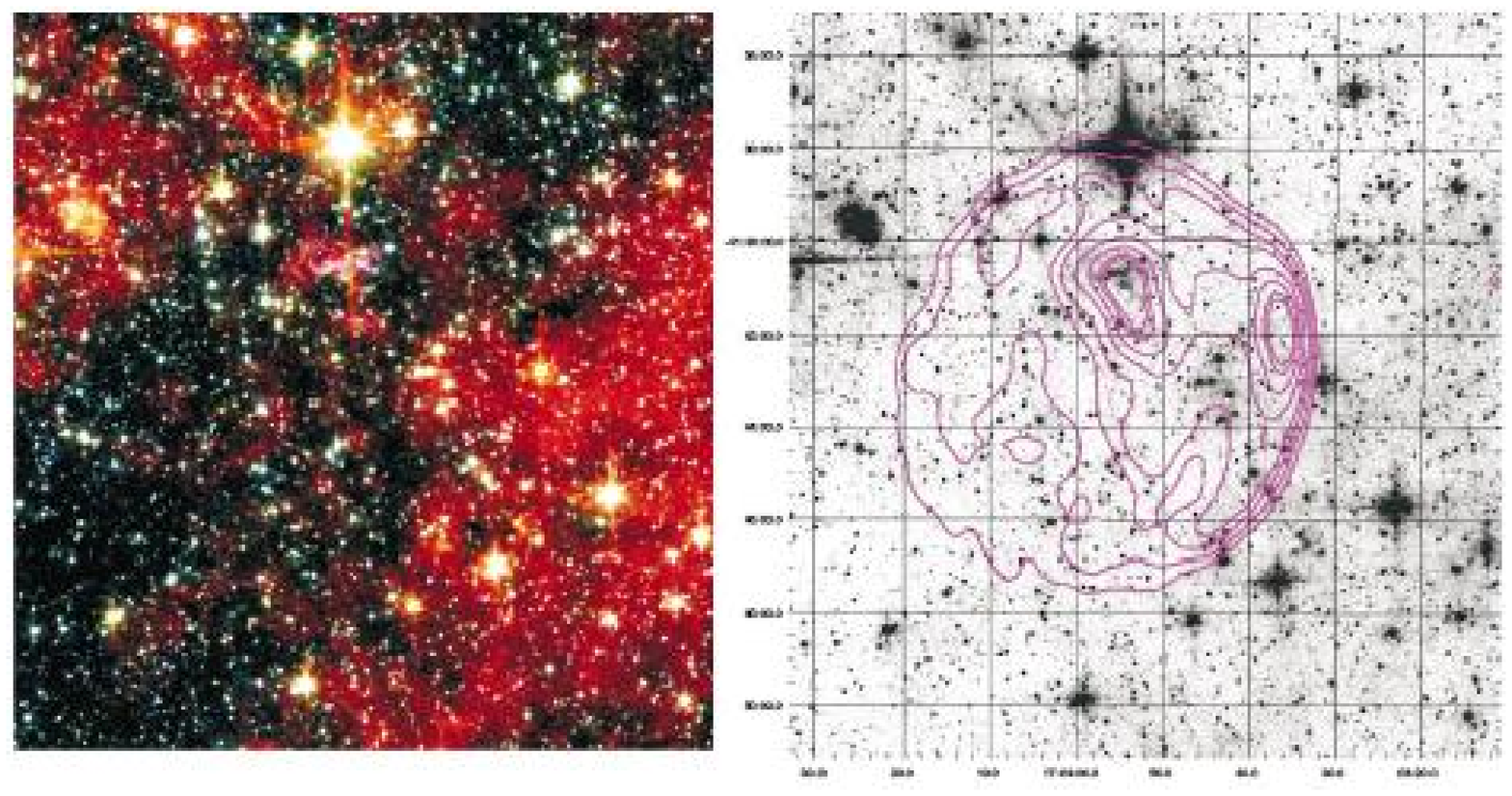}
\epsscale{1}
\figcaption[f18.eps]{
(left) {\it Spitzer}/IRAC color image of G344.7-0.1, with
red=channel 4, orange=channel 3, green=channel 2, blue=channel 1,
and magenta=channel 3 minus 0.3$\times$ channel 4.
The filament associated with the N radio peak in the SNR 
is relatively bright in channel 3 and appears `magenta' in this image.
(right) {\it Spitzer}/IRAC channel 3 image of G344.7-0.1, with
radio contours from the MSC overlaid. The greyscale ranges from 
15 to 30 MJy~sr$^{-1}$. The northern, interior radio peak 
has an apparently-associated mid-infrared filament.
{\it NOTE: this figure was degraded in quality for distribution on astro-ph.}
\label{G344.7image}}
\end{figure}

{\bf G344.7-0.1}---G344.7-0.1 has not been well studied, but
the radio morphology is shell-like with a bright northwestern shell 
\citep{dub93}.
Figure~\ref{G344.7image} 
reveals an area of irregularly-structured infrared emission
about $2'$ in diameter near \radec 17;03;55.1;-41;40;43. 
The colors (Table~\ref{detected}) suggest
the emission is likely due to shocked ionized gas.
Figure~\ref{G344.7image} shows the mid-infrared emission coinciding 
with the central radio peak from the MOST supernova remnant catalogue
\citep{MSC}. This is the first detection in the
infrared; it was not seen in previous IRAS surveys \citep{arendt,saken}.

\begin{figure}[th]
\plotone{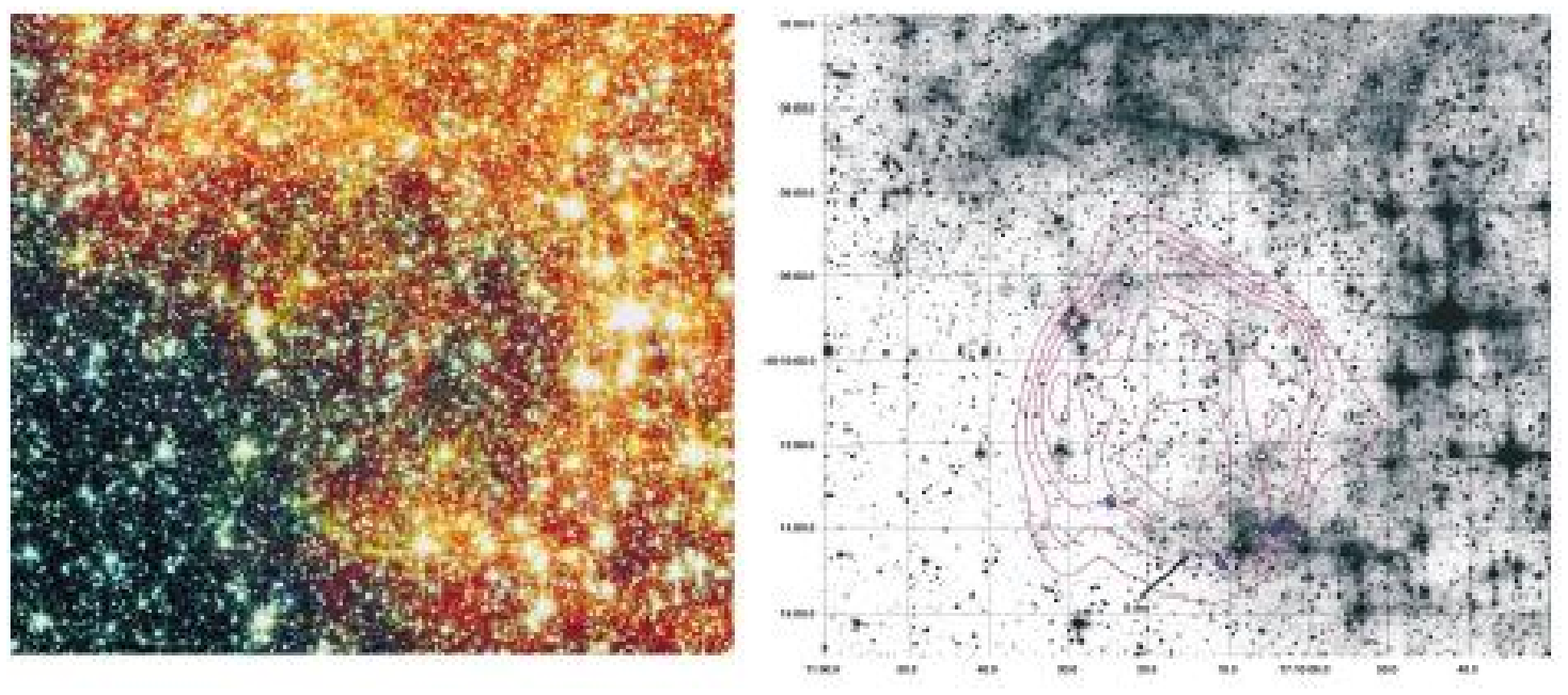}
\figcaption[f19.eps]{
{\it\bf (a)} {\it Spitzer}/IRAC color image of G346.6-0.2,
with red=8 $\mu$m, green=4.5 $\mu$m, and blue=3.6 $\mu$m. 
Radio contours from the MSC are overlaid in magenta;
contour levels are 50, 100, 150, 200, 250 mJy/beam in a
$43"\times 67"$ beam. 
{\it\bf (b)} {\it Spitzer}/IRAC 5.8 $\mu$m image of G346.6-0.2 with radio contours
(magenta) and the locations of OH 1720 MHz masers (blue diamonds) superposed.
The grescale ranges from 8 to 43 MJy~sr$^{-1}$.
{\it NOTE: this figure was degraded in quality for distribution on astro-ph.}
\label{G346.6combined}}
\end{figure}

{\bf G346.6-0.2}---The SNR is surrounded in the N and W by diffuse infrared emission
(5.8--8 $\mu$m). The emission associated with the remnant is a narrow
rim that follows the southern radio shell. The IRAC colors of the rim
are very distinct, so the rim is evident in the color image
(Figure~\ref{G346.6combined}a). 
This southern rim more-or-less connects the three
OH 1720 MHz masers associated with the remnant. 
The IRAC colors suggest molecular shocks, consistent with their close
association with the OH masers.
In addition to the southern rim, there 
is possibly some infrared emission following the northern shell,
but it is not readily evident in the Figures.

\clearpage

\begin{figure}[th]
\epsscale{.6}
\plotone{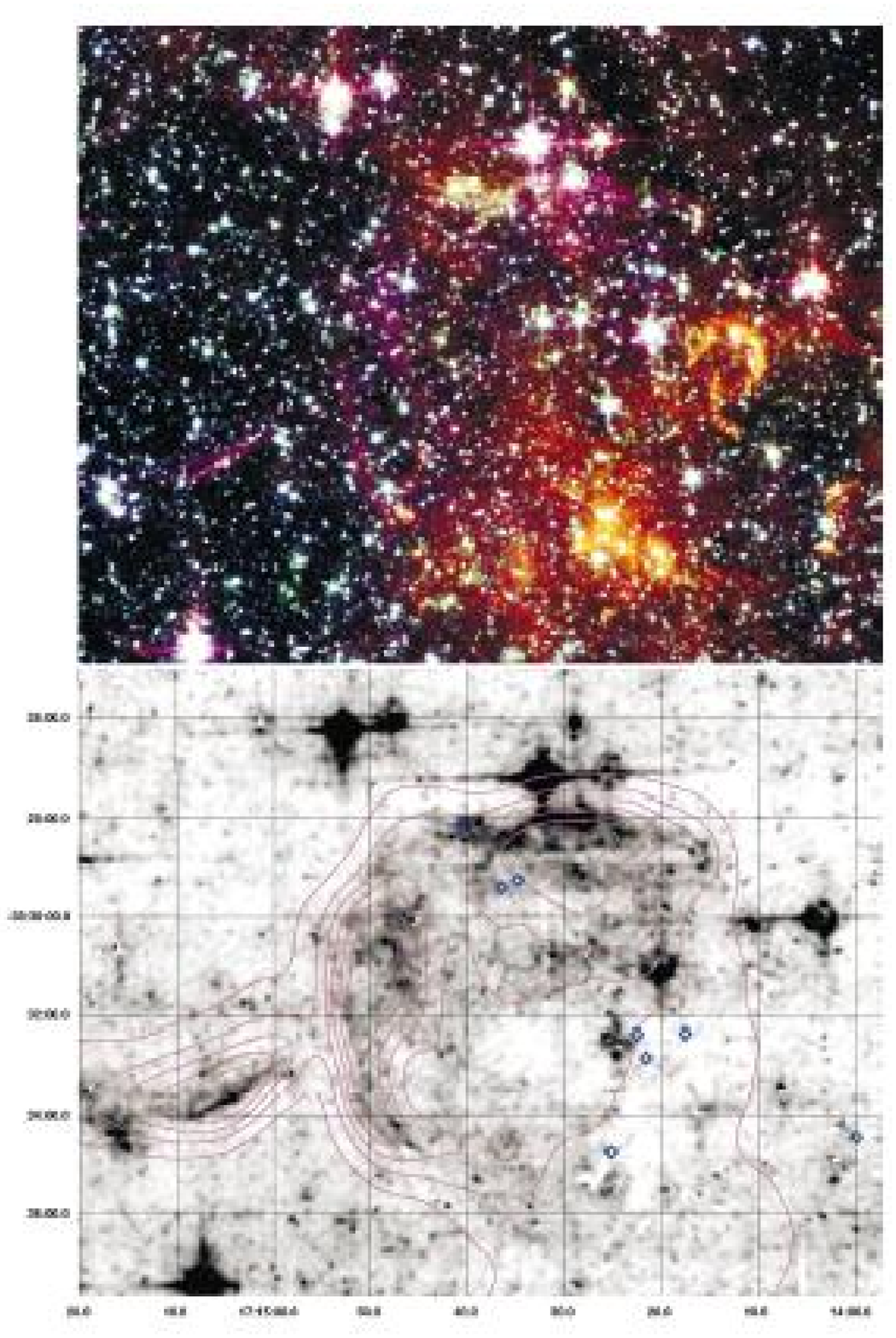}
\epsscale{1}
\figcaption[f20.eps]{\footnotesize
{\it\bf (a)} {\it Spitzer}/IRAC $[5.8]-[8.0]\times 0.361$
image of the supernova remnants CTB~37A (G348.5+0.1)
and G348.5-0.0. 
OH 1720 MHz maser positions \citep{frail96} are overlaid as diamonds.
No radio contours are overlaid, to avoid confusing the image,
in which diffuse emission is evident from both SNR.
CTB~37A is evident as a hemispheric shell including the relatively
bright northern arcs at \radec 17;14;35;-38;29;30 and western shell extending
through \radec 17;14;51;-38;31;50. G345.8-0.0 is evident as a short arc (labeled)
passing through \radec 17;15;05;-38;33;43.
{\it\bf (b)} {\it Spitzer}/IRAC color image of the supernova remnants CTB~37A (G348.5+0.1)
and G348.5-0.0,
with red=8 $\mu$m, green=4.5 $\mu$m, and blue=3.6 $\mu$m, 
magenta=$[5.8]-[8.0]\times 0.361$. 
The patch of green infrared emission in the NE is most likely shocked
molecular gas where the blast wave is impacting very dense gas. 
This green patch contains the northernmost OH 1720 MHz maser, which is at
-64.3 km~s$^{-1}$ \citep{frail96} like most of the other masers that are
associated with CTB 37A. 
G348.5-0.0 appears as a thin magenta filament (see Fig.~\ref{ctb37acombined} for
a guide), suggesting ionic shocks. 
The western shell of CTB~37A is not as easy to see in this
multi-wavelength image but has a magenta tint also suggesting ionic shocks.
{\it NOTE: this figure was degraded in quality for distribution on astro-ph.}
\label{ctb37acombined}}
\end{figure}

{\bf CTB~37A (G348.5+0.1)}---Figure~\ref{ctb37acombined}a shows
patches and filaments of  4.5 $\mu$m (green)
emission indicating shocked H$_2$ in the north,
as well as patches and filaments of 5.8--8 $\mu$m (red) emission
in the center and east of CTB~37A. 
The radio image has a `breakout' morphology suggesting impact into a denser 
medium in the NE and less dense in the SW. The 5.8$\mu$m-8$\mu$m image 
(Fig.~\ref{ctb37acombined}b) shows that the mid-infrared emission, probably from
shocked gas, has a similar morphology.
Green patches in the north correspond to the northernmost of the eight
OH 1720 MHz masers with velocities around -65 km~s$^{-1}$, which are
associated with CTB~37A. None of the other maser spots in
this remnant have associated 4.5 $\mu$m spots,
though some faint, patchy emission is in the vicinity of most of the
masers associated with CTB~37A.
The patch of green mid-infrared emission in the NE
is almost certainly shocked H$_2$ gas; the colors in IRAC channels
(Table~\ref{detected}) are similar to that expected for H$_2$+CO
and not consistent with PAH emission. The red filaments in the IRAC
image are clearly different, both in morphology and color. The
arc near \radec 17;14;14.7;-38;31;13 has IRAC colors 
\iraccolor 0.07;0.04;0.42;1, consistent with PAH emission. Since there are
so many regions with similar colors, it is possible that some of these
red filaments are unrelated to the remnant. We suspect, however, that
many of the infrared structures within the radio remnant are in fact
related to the remnant. For example, the relatively bright, semicircular
region including the filament mentioned above, the bright patch and
associated filaments centered around \radec 17;14;22.6;-38;35;06 
are all located
at the transition between the bright radio half-shell and the
fainter extended emission that appears to be the opening of the blowout, 
probably near the surface of the pre-explosion cloud. 
Long filaments, especially visible in the 5.8--8 $\mu$m images,
extend along the fainter NW and SW portions of the 
fainter, extended radio emission in the `blowout' shell, accurately 
delineating the boundary. The colors of these extended filaments are
consistent with PAH emission.

{\bf G348.5-0.0}---This remnant is partially superposed on CTB~37A;
high-resolution radio observations showed that G348.5-0.0 is a
separate remnant \citep{kassimCTB37}. The Spitzer/IRAC image shows a narrow
arc of emission that closely follows the radio rim. The correspondence
between the 5.8 $\mu$m filament detected by IRAC (Fig.~\ref{ctb37acombined}b) 
and the 6 cm radio rim detected with the VLA (Fig.~6 of Kassim et al. 1991)
is precise, so the association of these features is beyond doubt.
The filament is narrow---unresolved ($<3"$) to IRAC.
The filament is detected IRAC channels 2--4, with color ratios
(Table~\ref{detected}) measured toward \radec 17;15;04.6;-38;33;40.
The 5.8/8 $\mu$m ratio is much to high to be PAH emission, and the
filament is most likely dominated by emission lines from shocked gas.
The radio extent of this remnant can be seen in 
sensitive VLA images \citep{kassimCTB37}
to extend across the NW part of CTB~37A and
emerge from its northern rim. 
The second and third northernmost OH 1720 MHz masers in this region
are at $V_{LSR}=-23$ and -21 km~s$^{-1}$, clearly distinct from the velocities
of the other 8 masers that are associated with CTB~37A. There
are also distinct molecular clouds at these velocities that are
likely related to the remnants, with the -23 km~s$^{-1}$ cloud situated
just west of G348.5-0.0 having the masers at its edge; this cloud is
most likely associated with G348.5-0.0 \citep{reynosoCTB37}. 
The Spitzer/IRAC images further support the idea that G348.5-0.0 is
interacting with a dense cloud, because the infrared emission most
likely traces relatively dense, shocked gas.
The patch of green infrared
emission in the northern part of CTB~37A is in the region where that
remnant overlaps with G348.5-0.0 and could contain contributions
from both remnants, but we associate it with CTB~37A based on
the morphology (infrared filaments follow the CTB~37A radio shape, 
not G348.5-0.0, in this patch) and associated maser velocities.
There is no infrared emission
that appears related to the two OH masers associated with
G348.5-0.0. The infrared filament along the 
radio rim of G348.5-0.0 is, however, clearly part of this remnant, and 
projecting it to the west, taking into account the curvature of
the radio shell, it passes through the OH masers at -23 and -21
km~s$^{-1}$.
This remnant and CTB~37A are a special case of two different remnants
interacting with two different molecular clouds, all superposed in
the same several arcmin of the sky.

{\bf G349.7+0.2}---This is one of the brightest SNRs in the survey,
and owing to its great distance it is the most luminous.
Its X-ray brightness makes it one of the 
most luminous X-ray remnants and it may be younger than 3000 yr
\citep{slaneG349}.
The radio image was constructed by combining VLA data in A, C, CD, and D
configurations, with $5.0''\times 2.1''$ resolution \citep{brogan349}.
This remnant was shown to be interacting with a
large molecular shell by \citet{reynosoG349}, and OH maser
emission was detected by \citet{frail96}.  The mid-infrared emission is detected in
all IRAC bands (Fig.~\ref{g349color}); 
it is very bright at 5.8 $\mu$m and easily
discerned at 4.5 and 8 $\mu$m.
There are 4.5 and 5.8 $\mu$m emission peaks 
near 4 of the 5 OH 1720 MHz maser spots (all except the 
first in \citet{frail96}); these peaks are not as prominent at 8 $\mu$m.
Shocked H$_2$ line emission is a likely
contributor to the 4.5 and 5.8 $\mu$m wavebands. 
The 8 $\mu$m image may contain an emission mechanism in addition to 
the H$_2$S(5) line, such as [\ion{Ar}{2}] 6.99 $\mu$m. 
There is a filament of red (mostly 8 $\mu$m) 
emission extending from the supernova remnant
toward the west; it extends beyond the boundary of Figure~\ref{g349color}
and has colors similar to other extended emission in the field nearby.
This is apparently part of the dense cloud with which the supernova
remnant is interacting. The unusual morphology of the remnant, with
its shell much brighter on the western side, is caused by the 
shock propagating into the long axis of a roughly cylindrical cloud.

\def\extra{
\begin{figure}[th]
\plotone{../colorfigures/G349.7ch3lab.eps}
\epsscale{1}
\figcaption[../colorfigures/G349.7ch3lab.eps]{
{\it Spitzer}/IRAC 8.0 $\mu$m
image of the supernova remnants G349.7+0.1.
An elliptical outline of the radio image 
from the MSC are overlaid.
OH 1720 MHz maser positions \citep{frail96} are overlaid as diamonds.
\label{g349ch3}}
\end{figure}
}

\begin{figure}[th]
\plotone{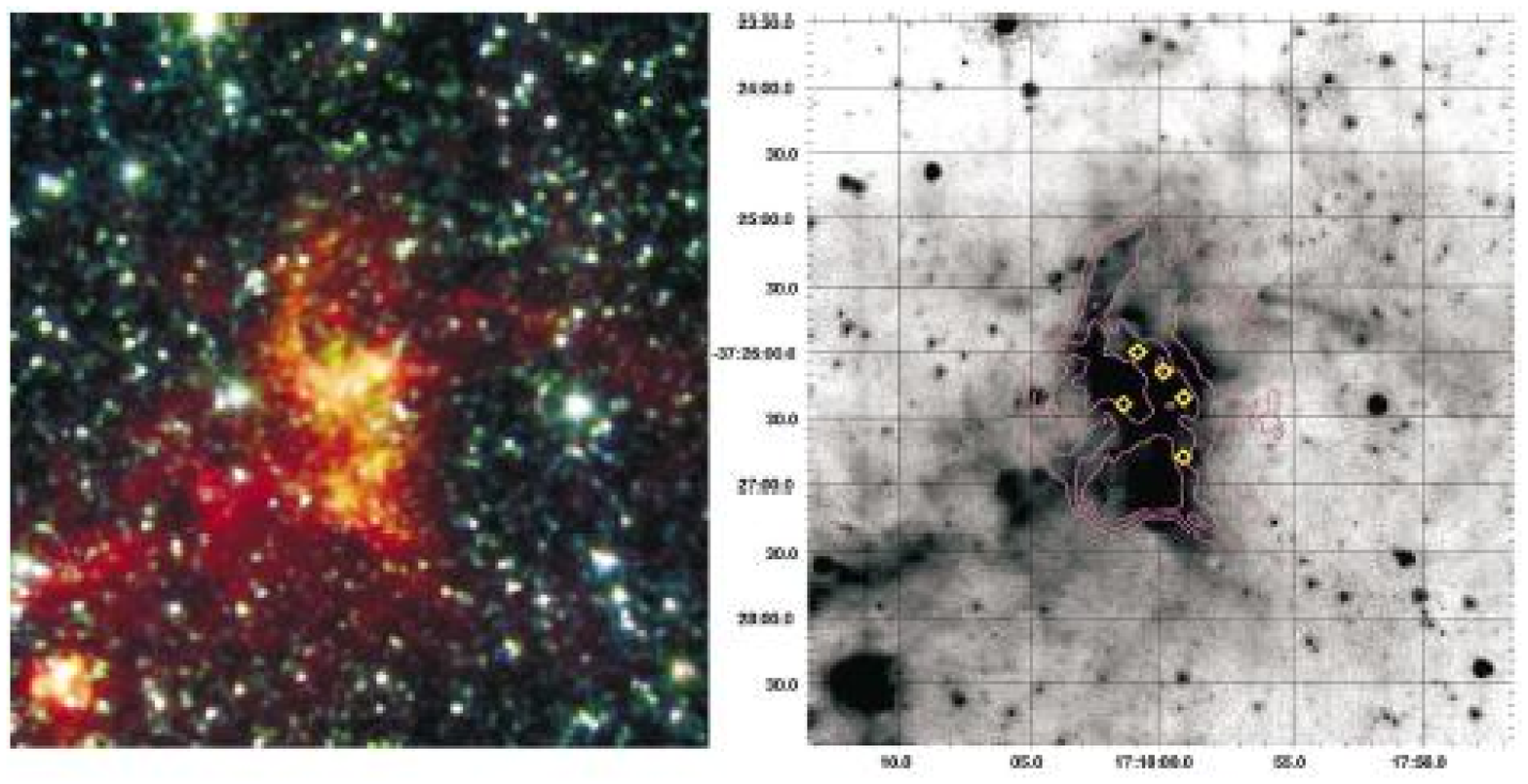}
\epsscale{1}
\figcaption[f21.eps]{
{\it Spitzer}/IRAC color image (left) and 5.8 $\mu$m image (right) of the 
supernova remnant G349.7+0.1.
For the color image, red=8 $\mu$m, green=4.5 $\mu$m, and blue=3.6 $\mu$m, 
magenta=$[5.8]-[8.0]\times 0.361$. Radio 20-cm contours are overlaid on
the channel 3 image.
{\it NOTE: this figure was degraded in quality for distribution on astro-ph.}
\label{g349color}}
\end{figure}
\clearpage


\section{Conclusions}

\begin{figure}[th]
\plotone{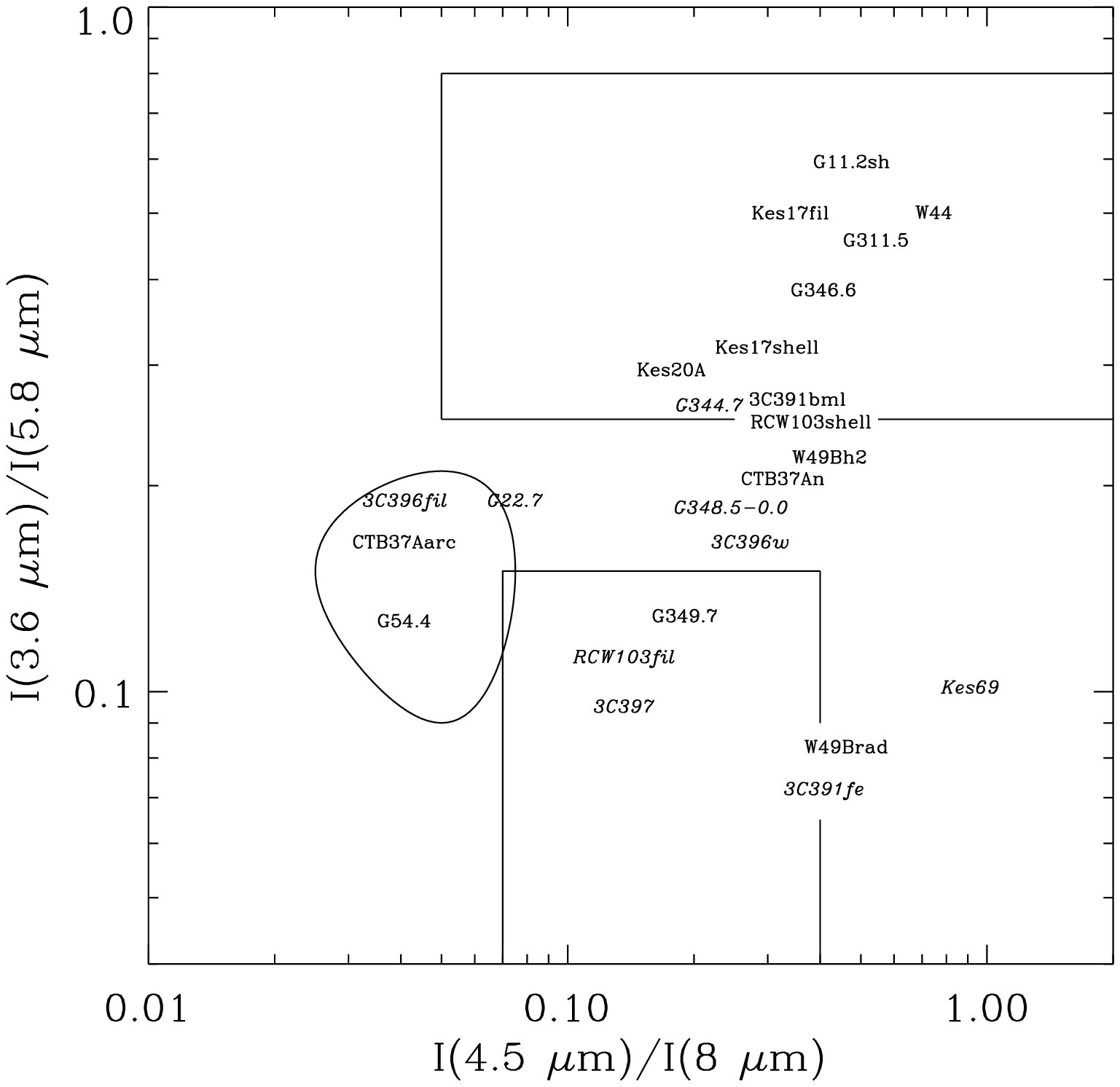}
\epsscale{1}
\figcaption[f22.eps]{IRAC color-color diagram for 1 or 2 locations in
each of the detected supernova remnants from this survey. The predicted
colors of molecular shocks (upper rectangle), ionic shocks (lower rectangle),
and PAH (circle) are copied from Figure~\ref{h2modsconcept} to aid interpretation.
Labels are in {\it italics} for remnants with upper limits (generally, these
are nondetections in channel 1 so the actual color would be {\it downward} from the
label in this figure).
\label{colcol}}
\end{figure}

The colors of the detected SNRs reveal a wide range of emission mechanisms.
Figure~\ref{colcol} shows the observed colors for 1-2 spots per remnant
together with bounding regions for molecular shocks, ionic shocks,
and PAH emission from Figure~\ref{h2modsconcept}.
Nine spots have colors consistent with molecular shocks,
\nion~ have colors consistent with ionic shocks, and ~\npah~ are
consistent with PAH emission from unshocked ISM. 
Of the remaining all but one are
intermediate between molecular and ionic shocks and probably represent
a mixture of shock types. One spot, the Kes~69 ridge, falls sufficiently 
outside the bounds of emission mechanisms considered here that its
colors cannot be explained by the mechanisms discussed here.

The SNRs with colors suggesting molecular shocks include 3C~391,
Kes~17, G346.6-0.2, G311.5-0.3, G344.7-0.1, Kes~20 and
G11.2-0.3. Two of these (3C~391 and G346.2-0.2) 
have associated OH 1720 MHz masers \citep{frail96,greenmaser}.
For comparison, only 14\% of the entire sample of 95 remnants have
OH 1720 MHz masers, so the association of these infrared colors with 
molecular shocks is not a chance coincidence. 
A total of 7 out of 18 IRAC-detected remnants have associated masers;
the 39\% association rate is again clearly not by chance. The remnants
with colors suggesting molecular shocks but lacking OH masers may
not have the specific, narrow physical conditions required to
generate the maser inversions. For example, 3C~391 has a
strong interaction with a giant molecular cloud on its bright radio NW
ridge, but the OH masers are only present in two spots to the edge of 
the main interaction \citep{rr02}.

One of the remnants with IRAC colors in the `molecular' region
of the color-color diagram is the historical SNR G11.2-0.3.
As discussed above, the mid-infrared emission from this young
SNR may not be exclusively from molecular lines.
Dust continuum is also unlikely to be the source
of the unusual colors, because it would make the 8 $\mu$m band bright and
move the colors to the left. Synchrotron radiation could contribute an
extra source of 3.6 $\mu$m and 4.5 $\mu$m emission and could possibly
explain the colors. Otherwise the colors could indicate
unusual abundances; further spectroscopic observations are needed
to understand this SNR.

The SNRs with colors suggesting ionic shocks are 3C~397, W~49B, and 3C~391 (NW ridge).
The radio emission from all three is bright and highly structured,
with bright near-infrared [\ion{Fe}{2}] that correlates in detail with 
the radio structure. 

The remnants with colors suggesting a mixture of molecular and ionic shocks
are RCW~103, CTB~37A, G348.5-0.0, and 3C~396. In RCW~103,
spectroscopy clearly reveals bright lines from both molecular and ionic shocks
\citep{olivaRCW103}. For CTB~37A and G348.5-0.0, OH 1720 MHz masers
are associated suggesting likely shocked molecular gas. The presence of
both types of shock in an SNR is not unexpected, as the shocked dense clumps that
cool via molecular lines are likely immersed in a lower-density medium
that, when shocked, cools via ionic lines. This is clearly seen in
the {\it ISO} spectra of W~28, 3C~391, W~44, IC~443, and RCW~103
\citep{rr00,olivaRCW103}.

\clearpage


The details of the emission mechanism for each SNR remain to be determined with
follow-up spectroscopy and comparison to ground-based observations. However, the
IRAC survey already suggests a trend such that mid-infrared-detected SNRs tend
to have colors suggesting shocks into dense gas. While this is an obvious selection
effect, because many important cooling lines of such shocks are present in the
mid-infrared, the sheer number of such detections suggests that molecular cloud 
interactions are not uncommon: at least 6\% of SNRs in our survey show infrared
colors suggesting molecular shocks. Some of these supernova remnants
suspected suspected to be interacting with molecular clouds have
been studied relatively little and await further investigation.

\acknowledgements  

This work is based in part on observations made with the {\it Spitzer Space
Telescope}, which is operated by the Jet Propulsion Laboratory, California
Institute of Technology under NASA contract 1407. We thank Rick Arendt for
his helpful comments.

\end{document}